\documentclass[12pt]{article}
\usepackage{graphicx} 
\usepackage{amsthm,amsmath,amsfonts,amssymb}
\usepackage[dvipsnames]{xcolor}
\usepackage{indentfirst}
\usepackage{bm}
\usepackage{float}
\usepackage{algorithm,algcompatible}
\usepackage{caption}
\usepackage{makecell}
\usepackage{mathtools}
\usepackage{multirow}
\usepackage{subcaption}
\newsavebox{\largestimage}
\usepackage[nodisplayskipstretch]{setspace}
\usepackage{microtype}
\usepackage{booktabs} 
\usepackage{textcomp}
\usepackage[hidelinks]{hyperref}\pdfoutput=1
\usepackage{hyperref}
\hypersetup{
  pdfinfo={
    Title={Your Title Here},
    Author={Your Name Here},
    Subject={If you want to put something here, do so},
    Keywords={Add some keywords if you feel so inclined}
  }
}
\usepackage{pdfpages}
\usepackage{natbib}
\usepackage{geometry}

\theoremstyle{plain}

\theoremstyle{definition}

\theoremstyle{remark}

\newcommand{\blind}{1}

\newcommand{\minitab}[2][l]{\begin{tabular}{#1}#2\end{tabular}}

\begin{document}

\def\spacingset#1{\renewcommand{\baselinestretch}%
{#1}\small\normalsize} \spacingset{1}


\if1\blind
{
  \title{\bf Solar Imaging Data Analytics: A Selective Overview of Challenges and Opportunities}
  \author{Yang Chen\thanks{All correspondence goes to Yang Chen: ychenang@umich.edu. 
    The authors gratefully acknowledge funding from NSF and NASA.}\hspace{.2cm}\\
    Department of Statistics and 
    Michigan Institute for Data Science\\ 
    University of Michigan, Ann Arbor\\
    and \\
    Ward Manchester \\
    Department of Climate and Space Sciences and Engineering\\
    University of Michigan, Ann Arbor\\
    and\\
    Meng Jin\\
    Lockheed Martin Solar and Astrophysics Laboratory\\
    and\\
    Alexei Pevtsov\\
    National Solar Observatory}
  \maketitle
} \fi

\if0\blind
{
  \bigskip
  \bigskip
  \bigskip
  \begin{center}
    {\LARGE\bf Title}
\end{center}
  \medskip
} \fi

\bigskip
\begin{abstract}
We give a gentle introduction to solar imaging data, focusing on the challenges and opportunities of data-driven approaches for solar eruptions. The various solar phenomena prediction problems that might benefit from statistical methods are presented. Available data prodcuts and softwares are described. State-of-the-art solar eruption forecasting models with data-driven approaches are summarized and discussed. Based on the characteristics of the datasets and state-of-the-art approaches, we point out several promising directions to explore from statistical modeling and computational perspectives. 
\end{abstract}

\noindent%
\vfill

\newpage
\spacingset{1.9} 

\section{Introduction}

\subsection{Introduction to Space Weather and {Various Solar Phenomena}}

\begin{figure}[tbph]
    \centering
    \includegraphics[width=0.8\textwidth]{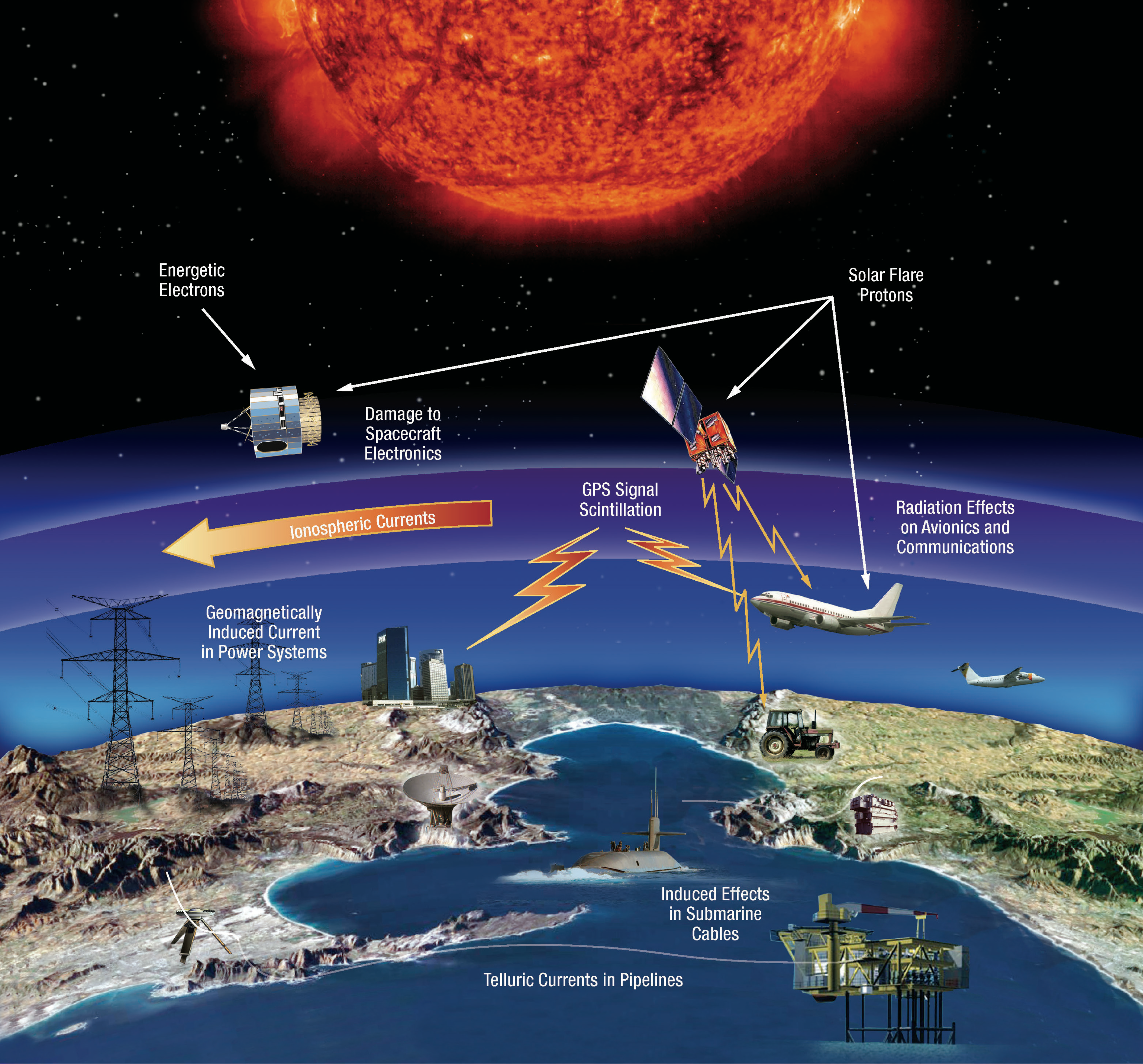}
    \caption{Technological and infrastructure affected by space weather events by the NASA Scientific Visualization Studio, \url{https://svs.gsfc.nasa.gov/31248/}.}
    \label{fig:swimpacts}
\end{figure}

{Broadly speaking, space weather refers to variable conditions in the solar system environment produced by the Sun's activity on relatively short time scales \citep[e.g.,][]{Baker1998,Schwenn2006}. Long-term effects in space weather on time scales longer than several solar rotations are referred to as space climate \citep[e.g.,][]{Mursula.etal2007}. The relation between the solar and geomagnetic activity (solar-terrestrial connections) was noted after \citet{Schwabe1843} discovered the 10-year sunspot cycle, when \citet{Sabin1852}, \citet{Wolf1852}, \citet{Gautier1852}, and Lamont \citep[see,][]{Reslhuber1852} reported on sunspot cycle periodicity in variations of Earth geomagnetic field. 
Significant variations in the magnetic field of Earth were noted concerning the first white-light solar flare, observed by \citet{Carrington1859} and \citet{Hodgson1859}. This flare caused one of the most extreme space weather events on record (the so-called Carrington event). Widespread sightings of aurora borealis, which lasted for about seven days and were observed as far south
as Cuba and Jamaica \citep[see a compilation of several observations in USA, Europe, and Asia, ][]{Silliman1859,Silliman1860b,Silliman1860c}. For a map of aurora sightings for this event, see \citet{Hayakawa.etal2019}.
The extended period of Aurora activity started a few days before the Carrington flare, which suggests that multiple eruptive events may have occurred during this period. The resulting geomagnetic activity had an impact on the telegraph, a global communication infrastructure of the time with the variable magnetic field induces electric currents in the telegraph wires strong enough to make them extremely hot or even spark fires \citep[see a compilation of several reports,][]{Silliman1860a}. 
The appearance of strong electric currents in telegraph wires in conjunction with aurora has been noted before by many observers \citep[e.g.,][]{Barlow1849,Silliman1860b}.
For a review of the early history of space weather impacts, including radio communications, navigation, radar detection, and artificial satellites, see \citet{Pevtsov2017} and visualization in Figure~\ref{fig:swimpacts}.
}

{Modern society’s well-being critically depends
on global technological systems. This includes global transmission networks of electric power, gas and
oil pipelines, communication and shipping logistics systems, and the global positioning system.
Early examples include power outages and blackouts \citep[][the most cited is the Hydro-Quebec
power grid collapse on 13 March 1989, which resulted in the loss of electric power to more than six million customers for nine hours]{Boteler.etal1998,Love.etal2022}, failure of communication satellites due to radiation damage by solar energetic particles (SEPs) associated with solar flares and coronal mass ejections (e.g., failure of Intelsat’s Galaxy-15 spacecraft in April 2010), electric discharge associated with a satellite-solar wind interaction originated from an elongated coronal hole \citep[Canadian Anik-E1 and E2 communication satellites,][]{Lam.etal2012}, or the enhanced atmospheric drag resulting from a major solar flare {or coronal mass ejection (CME)} event that caused a premature reentry of 38 out of 49 Starlink satellites launched by SpaceX \citep{Berger.etal2023}. In August 1972, a series of solar flares and associated geomagnetic storms led to widespread communications disturbances and power outages. These events include a CME with the shortest ever Sun–Earth transit time of just 14.6 hours, an average speed of 2850 km/s and a shock speed at 1 AU $>$ 1700 km/s \citep{Zastenker1978}. On 4 August, the geomagnetic storm caused a nearly instantaneous detonation of dozens of sea mines installed by the US during the Vietnam War south of Hai Phong \citep{Knipp.etal2018}. The 4 August flare, which occurred between the Apollo 16 (Apr.16-27, 1972) and Apollo 17 (Dec.  7-11, 1972) missions,
led to a significant increase in the radiation level in the interplanetary space. Later estimates \citep{Townsend.etal1991}
indicated that astronauts inside the spacecraft 
traveling to the Moon would have received radiation exposure that would exceed annual limits and approach career limits
for the skin and ocular lens. The average bone marrow
dose equivalent would have exceeded the recommended annual limit. For the astronauts outside the spacecraft (extravehicular activity, EVA, or on the moonwalk), the radiation dose equivalents would have been clinically significant, including nausea, vomiting, a very high probability of cataract formation, a slight increase in the probability of death.
Slightly smaller doses were estimated for a flare event in October 1989. 
For more detail on this topic of impact of SEP radiation on interplanetary travel see \citet[][and references therein]{Lockwood.Hapgood2007}.
Finally, in 1967, a large solar flare nearly triggered a nuclear war after it brought down the US early warning radar system with radio
noise, causing some military commanders to suspect Soviet jamming in preparation for a nuclear attack \citep{Knipp.etal2016}.
}

{Thus, it should not be surprising that space weather risk management has become a National priority.}
The threat-assessment report by the Lloyd’s Insurance company \citep{Lloyds:2013_chip_a} concludes that extreme events could cause 
\$0.6 -- 2.6 trillion in damage with a recovery time of months. An earlier report by the National Research Council \citep{Baker:2009a} arrived at similar conclusions. 
{More recent studies suggest that in case of an extreme solar and geomagnetic event, daily costs to the world economy would be several billion USD, not counting human costs. While extreme space weather events are rare, their damage could be catastrophic. Some studies estimate that the “economic costs associated with these extreme events have been heralded as being as high as \$1–2
trillion in the first year, equivalent to a so-called ‘global Hurricane Katrina’ ” \citep{Oughton.etal2017}. Due to the
global nature of space weather, the impact will affect a broad swath of the world economy, including power
distribution, transportation, communications, satellite infrastructure, aviation, and global positioning systems
\citep{Eastwood.etal2017}.  
}
Besides, penetrating radiation is one of the main obstacles to human exploration of Mars. 
 
{While the most dramatic impacts of space weather are due to
solar eruptive events, namely flares and CMEs, many other solar phenomena may also play an important role. For example, the coronal holes -- areas of open magnetic field, could be associated with recurrent geomagnetic storms of moderate intensity. The level of the solar UV and EUV radiation, which correlates with solar active regions, heats the Earth's upper atmosphere, increasing its density and increasing atmospheric drag on the Low Earth Orbit (LEO) satellites. Similarly, the increase in total electron content of the ionosphere affects the speed of radio wave propagation and, consequently, the accuracy of the global positioning system.  


{Here is a brief definition of solar phenomena, which are important for space weather:}
\begin{enumerate}
\item
{\textit{Solar flare} \citep[e.g.,][]{Bruzek.Durrant1977} is a sudden increase in electromagnetic radiation, typically in the range of UV to soft X-rays, from the solar atmosphere (mostly the chromosphere and corona). This radiation is primarily thermal in nature, coming from regions of localized heating caused by magnetic energy release. Acceleration of charged particles in flaring regions can also produce non-thermal radiation, such as hard X-rays. In the visible light (e.g., H$\alpha$ spectral line, $\lambda$ 656.3 nm, the flare importance is defined by its relative brightness (F/faint, N/normal, and B/bright) and total area (S/subclass, 1-4 classes). In X-ray, the flares are characterized by the maximum intensity in the 0.1-0.8 nm wavelength band: 
\begin{center}
\begin{tabular}{cc}
\hline
Class & Maximum Intensity Range\\
\hline
A & (1-9) $\times$ 10$^{-5}$ erg/(cm$^2\cdot$ s)\\
B & (1-9) $\times$ 10$^{-4}$ erg/(cm$^2\cdot$ s)\\
C & (1-9) $\times$ 10$^{-3}$ erg/(cm$^2\cdot$ s)\\
M & (1-9) $\times$ 10$^{-2}$ erg/(cm$^2\cdot$ s)\\
X & (1-9) $\times$ 10$^{-1}$ erg/(cm$^2\cdot$ s)\\
\hline
\end{tabular}
\end{center}
}
The A{-class} flares are near the background energy level. 
{A- and B-class flares are typically not geoeffective. C-class flares may have some geomagnetic impact, and M- and X-class flares are the major flares with a major space weather impact. Major flares identified as Importance 3 and 4 in visible light observations typically correspond to X-ray class M and X. \citet{Hayakawa.etal2020} 
find that optical flares with an importance of 3 were associated mainly with X-class (66\%) and M-class (30\%) flares. In 4\% cases, strong H$\alpha$ flares with C-class flares.
}
\item
{\textit{Coronal Mass Ejection} \citep[CME,][]{Webb.Howard2012} is a large structure of magnetized plasma expelled from the Sun into the interplanetary space. They are associated with strong flares (C, M, and X classes) and eruption of the chromospheric filaments (see Figure \ref{fig:filamentCME}). Their geoeffectiveness depends on their speed \citep{Srivastava.Venkatakrishnan2002} and the orientation of the magnetic field. Those situated west of the solar central meridian, with high velocity (fast) and a southward orientation of the magnetic field, are more geoeffective, with magnetic field orientation being the most important parameter. Some CMEs may not have a clearly identifiable source region \citep[so called, stealth CMEs,][]{Ma.etal2010}. They could be associated with the eruption of magnetic structures of filament channels without significant filament material in it \citep[e.g.,][]{Pevtsov.etal2012}.
}
\item
{\textit{Chromospheric filament} \citep[prominance at the limb,][]{Gibson2018} is a ``cloud'' of cooler and denser plasma suspended in the chromosphere and corona by the magnetic field. The vast majority of CMEs are associated with filament channels, and early filament rise was found to proceed the CME initiation \citep[see,][and references therein]{Berezin.etal2023}. The orientation of the magnetic field and its handedness (chirality, or sign of magnetic helicity) are found to provide a good representation of the southward orientation of the magnetic field in CMEs. Filaments are located along the polarity inversion line of large-scale magnetic fields at the photosphere.
}
\item
{\textit{Coronal hole} \citep[CH,][]{Cranmer2009} can be identified in coronal images as the darkest (lowest intensity) areas on the solar disk, which can persist for several solar rotations. Historically, CHs have been identified using observations in He I 1083.0 nm (near infrared), although their appearance is not as evident as in extreme UV (EUV) or X-ray images.
CHs are associated with weak, mostly unipolar magnetic fields opened to the interplanetary space. CHs are the source of fast solar wind. Because CHs may exist for several solar rotations, they are associated with the recurrent geomagnetic storms occurring on Earth with the period of solar synodic rotation (about 27.2753 days). The geomagnetic storms associated with CHs are typically moderate in strength. Perhaps the largest space weather risk from fast-speed streams originating from the coronal holes comes from satellite charging. The resulting electrostatic discharge can damage electronics, phantom commands, cause loss of operations, and in some cases, total satellite loss \citep[][]{Horne.etal2018}.
}
\item
{\textit{Solar Energetic Particles} \cite[SEPs,][]{Reames2021}.
SEPs are the radiation events associated with high-energy particles originating from the Sun in the energy range from about 10 KeV (kilo electron volts) to 10 GeV, lasting from hours to days. Their composition is mostly electrons and protons, but heavy elements (He through Au/Pb) have been identified as well. SEP events are classified as impulsive or gradual. The former originates as the result of magnetic reconnection in the solar corona at the location of flares, while the latter is the result of particle acceleration in the CME-driven shocks. Gradual SEPs are also long-duration events. Some SEPs have sufficient energy to cause localized increases in the radiation at the Earth's surface, so-called Ground-Level Events (or Ground-Level Enhancements, GLEs). 
}
\end{enumerate}

\begin{figure}[tbph]
    \centering
    \includegraphics[width=0.9\textwidth]{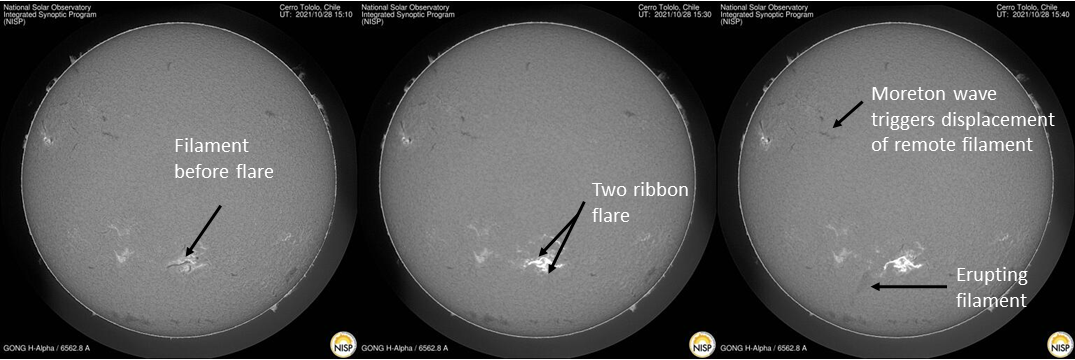}
    \caption{Two ribbon flare and filament eruption associated with CME and X1.0 X-ray class flare as observed by GONG/CT on 28 October 2021. Black arrows mark the approximate position of the filament before and during its eruption, bright flare ribbons, and a signature of the Moreton (blast) wave triggered by this major flare. Off-limb, halo is due to uncorrected scattered light in the H$\alpha$ filter. Bright features extending beyond the solar limb are prominences, which appear as dark features when they are observed on the solar disk. Source: NSO FY 2021 Annual Progress Report and FY 2022 Program Plan
    \url{https://nso1.b-cdn.net/wp-content/uploads/2023/03/NSO-2022_23_Final.pdf}}
    \label{fig:filamentCME}
\end{figure}

\subsection{State-of-Art Solar Events Forecasting with Physics}

State-of-the-art solar event forecasting has been based on physical and empirical models. {Empirical solar flare prediction models primarily focus on parameterizing relationships between the active region's photospheric magnetic field (e.g., magnetic topology, Lorentz force, free energy, helicity, etc.) and identifying relationships between these parameters and solar flare activity~\citep[e.g.,][]{McIntosh90,Falconer2002, leka2003photospheric,schrijver2007characteristic,fisher2012global,moore2012limit}. 
The state-of-the-art in physics-based solar flare prediction involves realistic MHD modeling of active region evolution by driving the model's inner boundary with time-series magnetic field measurements, allowing for the self-consistent modeling of the pre-flare energy build-up process~\citep[e.g.,][]{cheung2012method,jiang2016ncomms}. 
However, due to the limitations in both physical realism, computational cost, and observational data, this advanced modeling remains at the research level. It is not yet applicable for operational solar flare prediction.}

{Although CMEs and flares are both magnetically driven events, with a correlation that grows with energy, there is no strict one-to-one correspondence between them. To understand the physical mechanism distinguishing flares and CMEs, significant effort has been devoted to analyzing the structural properties of the global coronal magnetic fields, which may play a crucial role in determining whether an eruption evolves into a CME or remains a confined flare~\citep[e.g.,][]{torok2005confined,devore2008homologous,liu2008apjl,baumgartner2018factors}.
The CME prediction model uses a similar approach to the solar flare prediction model, aiming to find relationships between CME productivity and the features of the photospheric magnetic field~\citep[e.g.,][]{qahwaji2008automated,bobra2016predicting,kontogiannis2019photospheric}.

{Physics-based SEP prediction models aim to numerically incorporate the relevant physics of particle acceleration, encompassing the background solar corona and wind environment, CME eruptions and their driven shocks, particle acceleration, and transport. Each of these processes is an active research area and not fully understood, making SEP prediction extremely challenging. For details about the current state-of-the-art SEP prediction models, refer to the review paper by \citet{whitman2022review}.}

\subsection{Promises of Data-Driven Approaches}

The current space weather forecasting based on physical models is far from reliable: the forecasting window is only minutes, and the accuracy is low. Recently, with much more data becoming available, data-driven approaches are gaining attention in the space science community; see~\citet{Leka:2018b} and~\citet{Camporeale:2019a} for a review. How to make the best use of the large amount of data available to provide reliable real-time forecasting of space weather events is one of the major questions for scientists in the field. Recent work published in the Space Weather journal~\citep{bobra2015solar,nishizuka2018deep,chen2019identifying,wang2020predicting,jiao2020solar,nishizuka2021operational} shows that it is highly promising to extend the solar flare forecast time scale from minutes to days using machine learning methods. These results are highly encouraging for the field and provide a benchmark for future studies. The authors of this paper have published multiple papers on adopting existing ML methods and developing novel statistical methods for solar flare forecasting in the past few years. More specifically, \citet{chen2019identifying,wang2020predicting} adopt the Long Short Term Memory (LSTM) neural network to classify strong solar flare events, demonstrating that constructing precursors of solar flare events from our set of predictors is feasible. In \citet{jiao2020solar}, we combined the idea of a mixture model and the LSTM to propose a mixed LSTM regression model that predicts the flaring label and flare intensity jointly in one single optimization problem. This results in improved performance in strong flare forecasting. In \citet{sun2021improved}, we adapt ideas from spatial statistics and topological data analysis to construct physically interpretable predictors of solar flare events, further improving strong flare forecasting performances. In \citet{sun2022predicting,mm2022flare}, we explore the potential of combining heterogeneous sources of data (data from different instruments from two different solar cycles) for solar flare forecasting, thus seeking to mediate the sample size limitation in rare events (strong solar flare) prediction. In \citet{sun2023tensor}, we propose a new tensor regression model that combines information from multiple sources of data, accommodating their spatial, temporal, and spectral continuity and sparsity, and propose an efficient computational algorithm to solve the problem.

\subsection{Challenges for Solar Flare Forecasting}

The complexity of solar flares and the \textit{infrequent} occurrence of energetic events results in highly heterogeneous data with large variability, making fast and accurate predictions of the time and intensity multiple hours/days ahead an extremely challenging task. According to the \citet{NOAAscale}, during solar cycle 24 (roughly 11 years), there were  $>2000$ M flares and less than $180$ X flares. Figure~\ref{fig:solarcycle} shows the sunspot number, which characterizes the solar activity levels of multiple solar cycles. We can see that the different phases of the solar cycle demonstrate significantly different levels of solar activities. Our high-quality observational data to be described in the next section typically only cover one or a little more than one solar cycle. 

\begin{figure}
    \centering
    \includegraphics[width=0.9\textwidth]{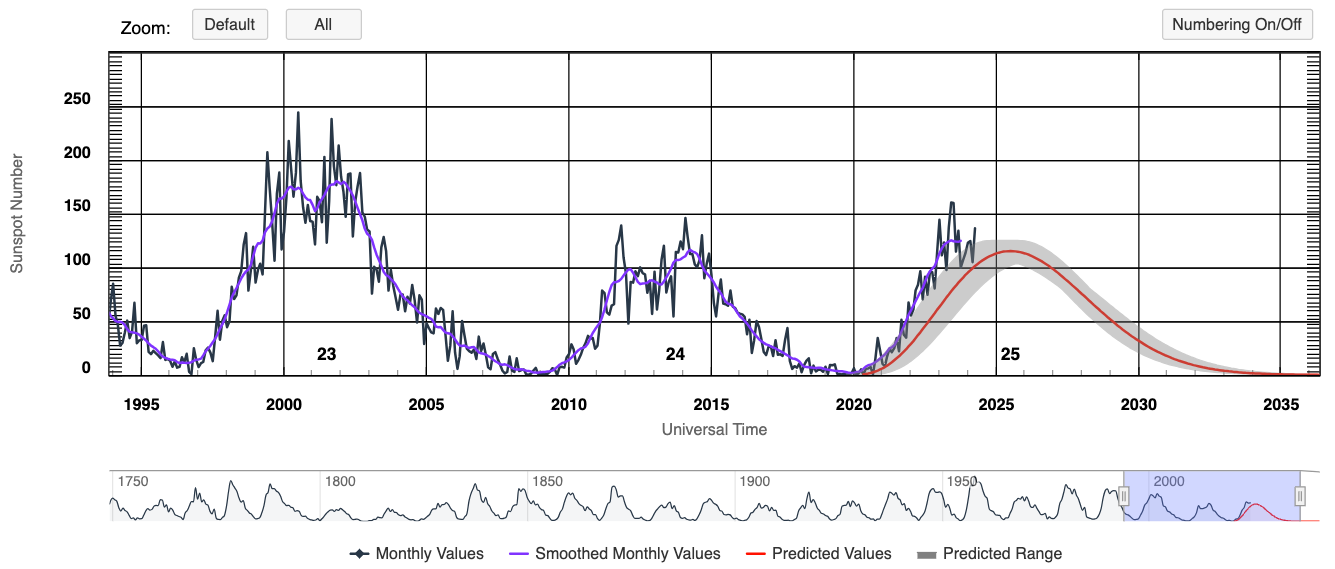}
    \caption{Solar cycle progression from NOAA/SWPC, \url{https://www.swpc.noaa.gov/products/solar-cycle-progression}.}
    \label{fig:solarcycle}
\end{figure}

Moreover, since A and B class flares are relatively weak and close to background levels, many of them are not recorded, causing significant ``missing data''. Off-the-shelf machine learning algorithms cannot achieve desirable performance; thus, more delicate modeling is required to carefully extract the maximum amount of information from the complex structured data. What exacerbates the situation for data-driven methods, especially complex modeling, is the computational cost required to process the high-resolution and high-cadence observations over an extended period. We will detail these aspects when introducing solar data sets in section~\ref{sec:data}.

Scientifically, it is important to (a) identify the solar active regions that have high potential to erupt in an automated fashion, (b) extract features from observed solar images using principled statistical algorithms, (c) most importantly, provide a probabilistic forecast for the eruption time, magnitude and magnetic field configuration, and (d) facilitate new understandings of the mechanism/physics of space weather events. {To achieve these goals, we need (1) careful statistical modeling of the spatial and temporal patterns, thus identifying precursors that finally do (or do not) lead to rare and extreme events, and (2) advanced computational techniques to handle the massive and multi-faceted data sets. }

\section{Massive Solar Imaging Data}

\label{sec:data}

There is a broad spectrum of data given by various observatories, and a multitude of information is covered about solar activities. Figure~\ref{fig:earthsun} shows the Sun-Earth interactions and layers of the solar and near-Earth environments. Figure~\ref{fig:obs} shows the NASA Heliophysics Systems Observatory located near-Earth or near-Sun. The data features, quality, and accessibility vary across observatories. In this paper, we focus on solar imaging data, which are typically given by the Flexible Image Transport System (FITS), i.e., as multidimensional numerical arrays. This section introduces potential available data sources and gives concrete processed data that machine learning researchers have adopted. 

\begin{figure}[tbph]
    \centering
    \includegraphics[width=0.9\textwidth]{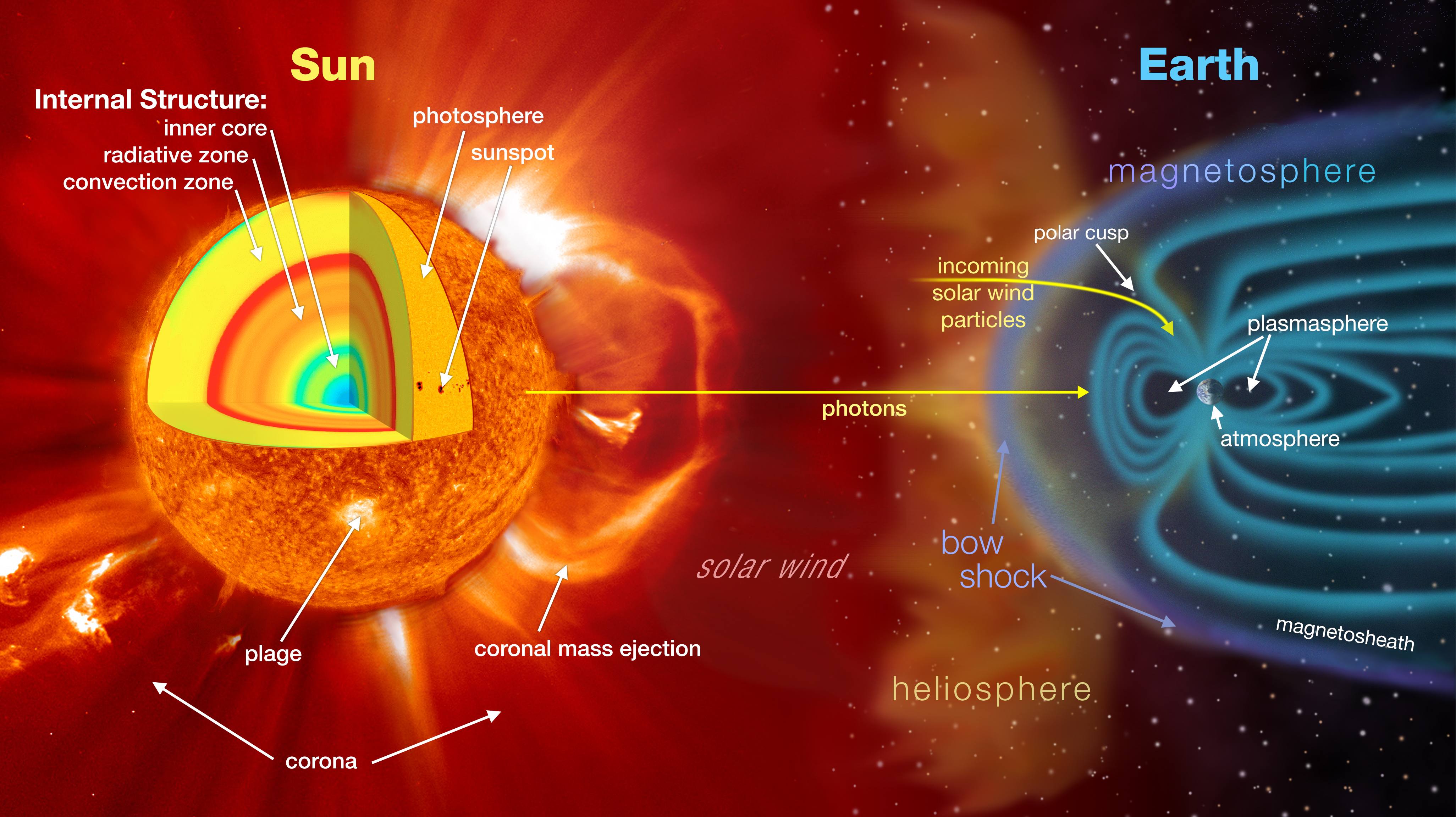}
    \caption{Illustration depicts Sun-Earth interactions that influence space weather. Source from NASA SVC: \url{https://svs.gsfc.nasa.gov/30481/}. }
    \label{fig:earthsun}
\end{figure}

\begin{figure}[tbph]
    \centering
    \includegraphics[width=0.9\textwidth]{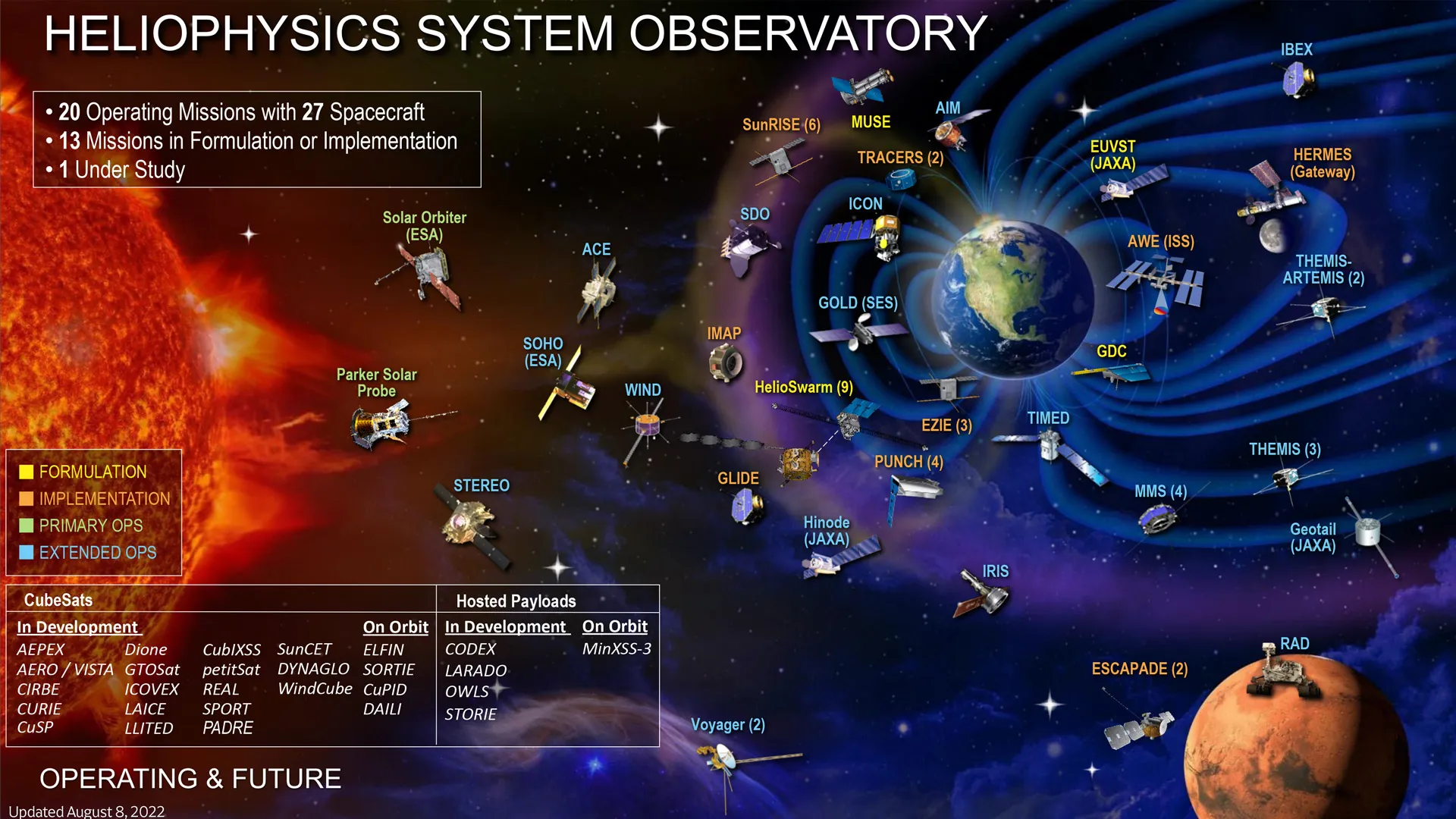}
    \caption{``The NASA Heliophysics Systems Observatory works with other NASA systems observatories to give NASA a complete picture of the Sun-Earth System. '', figure and words are taken from \url{https://science.nasa.gov/learn/heat/missions/}.}
    \label{fig:obs}
\end{figure}

\subsection{Overview of Available Data Sources}
\label{subsec:overview_data}

As shown in Figures~\ref{fig:earthsun} and~\ref{fig:obs}, there is a broad spectrum of observatories taking measurements about the Sun and the near-Earth environment, providing thorough information about the Sun-Earth interactions. 
Here, we review three major types of solar observations relevant to monitoring and forecasting solar activities. These observations are the ones most popularly adopted by scientists hoping to use data-driven approaches for space-weather monitoring. 
{The most relevant are the imaging data, which depict solar evolution in the solar photosphere (visible ``surface'' of the Sun), chromosphere (the portion of solar atmosphere above the photosphere), and corona (the outer part of the solar atmosphere). Imaging data may originate from spaceborne and ground-based instruments. Most imaging data are taken at or near Earth and provide the Sun's view from Sun-Earth vantage (or direction). The most notable examples include GONG, SDO, SOHO, SOLIS, {STEREO (which has been on the far side of the Sun from Earth)}, GOES, Hinode, and Yohkoh (see, Table \ref{tab:observatries}).
The direct imaging data are used to (i) identify and classify solar features (sunspots and active regions, coronal holes, filaments, CMEs, etc.) and (2) to create catalogs of events (e.g., solar flare catalog by NOAA, Yohkoh, and Hinode; and SOHO LASCO CME catalogs, CDAW (\url{https://cdaw.gsfc.nasa.gov/CME_list/index.html}) and Cactus (\url{https://www.sidc.be/cactus/}). GOES X-ray flux records are used to identify the X-ray flares and their properties.
}

\begin{table}[tbh]
\begin{minipage}{\textwidth}
    \centering
\begin{tabular}{|c|c|c|c|c|}
    \hline
    Observatory & Location & Type of data & Years & Link\\
        \hline
        GONG\footnote{NSF's Global Oscillation Network Group} & Ground-based &LOS magnetograms, &1995-P & \url{www.gong.nso.edu}\\
        \hline
        & &H$\alpha$ images, helioseismology& &\\
        SOLIS\footnote{NSF's Synoptic Optical Long-term Investigations of the Sun} & Ground-based & Vector/LOS magnetograms, & 2003-2017 & \url{solis.nso.edu}\\
         & &H$\alpha$, He 1083.0 hm images& &\\
         \hline
         SDO\footnote{NASA's Solar Dynamics Observatory}&Geocentric& Vector/LOS magnetograms, & 2010-P & \url{sdo.gsfc.nasa.gov}\\
                 &orbit&EUV images, helioseismology& &\\
        \hline
        STEREO\footnote{NASA's Solar-Terrestrial Relations Observatory}&1 AU orbit& EUV images, \textit{in situ}&2006-P&\url{stereo.gsfc.nasa.gov}\\
         & &(SEPs, magnetic field)& &\\
         \hline
         GOES\footnote{NOAA's Geostationary Operational Environment Satellite}&GEO&X-ray flux, EUV images, &1975-P&\url{www.swpc.noaa.gov}\\
         & &SEPs& &\\
         \hline
         SOHO\footnote{ESA-NASA's Solar and Heliospheric Observatory($^*$MDI stopped observing in 2011)} &Sun-Earth &LOS magnetograms, & 1996-P$^*$&\url{soho.nascom.nasa.gov}\\
         & Lagrange L1 &EUV images, coronagraph& &\\
         \hline
         Yohkoh\footnote{Japan's solar observatory spacecraft}&LEO&soft and hard X-ray&1991-2001&\url{umbra.nascom.nasa.gov/}\\
         &&&&\url{yohkoh/y4sdac_top.html}\\
         \hline
         Hinode\footnote{Japan/UK/US mission}&LEO&vector magnetograms&2006-P&\url{science.nasa.gov/}\\
         &&EUV imaging, soft X-ray&&\url{mission/hinode}\\
        \hline
    \end{tabular}
    \end{minipage}
    \caption{Major Heliophysics Observatories with imaging data. LEO (Low Earth Orbit): an orbit relatively close to Earth’s surface. GEO (Geostationary orbit): a circular orbit 22,236 miles above Earth's Equator, a satellite's orbital period is equal to Earth's rotation period of 23 hours and 56 minutes. Geocentric orbit: objects orbiting Earth, such as the Moon or artificial satellites.}
    \label{tab:observatries}
\end{table}

GONG is a global network of six solar robotic telescopes located in Australia, India, the Canary Islands (Spain), Chile, and the USA (California and Hawai'i). GONG provides full disk images of the solar photosphere (broadband or white light) and chromosphere (H$\alpha$), line-of-sight (LOS) magnetograms, and Dopplergrams (helioseismology data) in the photosphere. Helioseismology data are used to derive the so-called farside imaging (low-resolution maps of solar magnetic fields on the side of the Sun opposite to one facing the Earth). The Michelson Doppler Imager (MDI) on board SOHO also produced LOS magnetograms and Doppler maps similar to GONG, albeit with a lower spatial resolution. The Extreme Ultra-violet Imaging Telescope (EIT) and the Large Angle and Spectrometric Coronagraph (LASCO) are other instruments on board SOHO spacecraft providing solar corona images. The Helioseismic and Magnetic Imager (HMI) measures 3D magnetograms, and the Atmospheric Imaging Assembly (AIA) photographs the Sun's atmosphere and the corona in multiple wavelengths.


Next, we describe H-$\alpha$ images by GONG and other observatories in section~\ref{subsec:Halpha}, GOES observations (including flare list and X-ray flux measurements) in section~\ref{subsec:GOES}, SDO observations in section~\ref{subsec:SDO}, SOHO and GONG magnetogram observations in section~\ref{subsec:SOHOGONG}.

\subsection{H-alpha Images: GONG and others}
\label{subsec:Halpha}

{From the observation perspective, full disk H$\alpha$ (Hydrogen–$\alpha$) images can be used to identify the chromospheric filaments, solar flares, and their properties in the visible wavelength range. H$\alpha$ is one of the commonly used spectral lines in solar astronomy, and thus, there are a number of ground-based instruments that provide such observations. 

As some examples, one can mention the Chromospheric Telescope  \citep[ChroTel, ][years of operations 2012--2020]{Kentischer2008, Bethge2011}, H$\alpha$ instruments at the Global Oscillation Network Group \citep[GONG, ][2010--present]{Harvey1996,Hill2018,NISPHA2010}, the Kanzelh\"ohe Observatory \citep[KSO, ][]{Otruba2003, Poetzi2015, Poetzi2021}. For additional details of these instruments and their application in machine learning, see, e.g., \citet{Diercke.etal2024}.
ChroTel and H$\alpha$ instruments at KSO are single-site instruments, and thus, their observations are limited to a day-night cycle, weather permitting. GONG is a 6-site global network, which typically provides a 90\% duty cycle (90\% of the 24-hour period is covered by observations). Each GONG site takes H$\alpha$ observation every 60 seconds. See Figure~\ref{fig:gonghalpha} for a snapshot of GONG H-$\alpha$ data updated by the NSO website on May 3, 2024. The time of observations at the adjacent sites is shifted by 20 seconds, which allows for the achievement of a network cadence of 20 seconds. The GONG data is used by the NOAA Space Weather Prediction Center (SWPC), the US Air Force 557th Weather Wing, and the NASA Community Coordinated Modeling Center (CCMC) to monitor space weather conditions. 

\begin{figure}[tbph]
    \centering
    \includegraphics[width=0.8\textwidth]{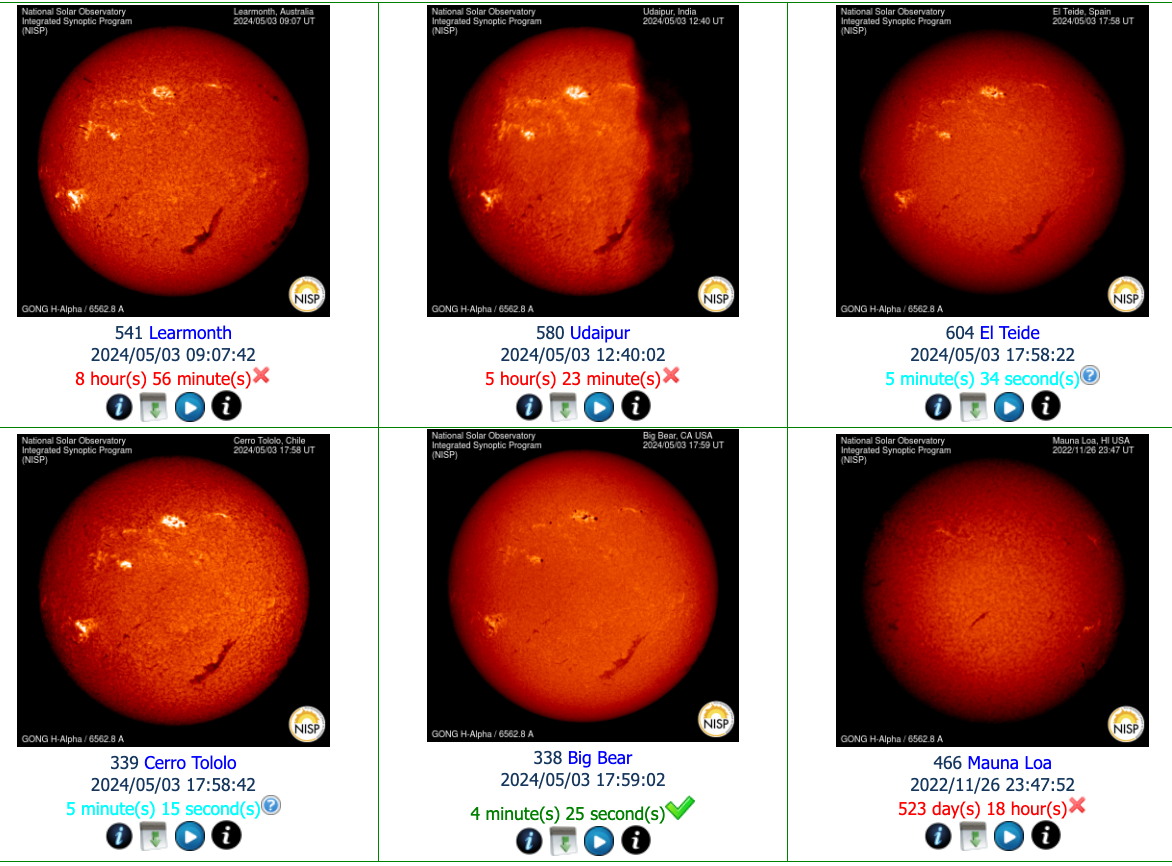}
    \caption{Summary of GONG H-$\alpha$ data from 3 May 2024 showing the latest observations from six network sites located at (upper panel, left to right) Learmonth Solar Observatory, Australia; Udaipur Solar Observatory, India, Teide Observatory, Canary Islands and (lower panel) the Cerro Tololo Inter-American Observatory, Chile; Big Bear Solar Observatory, California; and Mauna Loa Observatory, Hawai'i. The image from Udaipur is affected by a cloud, which partially covers the solar disk. Notations below each image indicate the date and time of observations. The last line under each image shows the time since the last image was taken. For this line, the text in red is used for sites that are not observing (due to weather or a night). The green text shows the sites currently observing, and the sites that are observing but have not returned a recent image (e.g., due to clouds) are shown in cyan.  GONG site at Mauna Loa Observatory was down since 26 Nov. 2022 after the volcanic eruption, when the lava flows shut down the operations. Thus, the image of the Sun from Mauna Loa is outdated and thus looks very different from the other sites. The dark elongated feature shown in 5 images is a quiescent filament. The bright, compact areas on the images correspond to three active regions. Larger areas, which appear slightly brighter than the surrounding background, correspond to the magnetic field of decaying active regions.
    The appearance of active regions and filaments can vary depending on the properties of the H$\alpha$ filter (c.f. image in the low-middle panel with low-left or upper panel). This difference in appearance may complicate the identification of solar active regions using the ML approach. This summary page is available at \url{https://gong2.nso.edu/products/tableView/table.php?configFile=configs/hAlphaColor.cfg}. Due to the dynamic nature of this page, images of different dates and times will be shown. All images are oriented with the solar North up and East to the left).}            
    \label{fig:gonghalpha}
\end{figure}

KSO H-$\alpha$ instrument is part of the Global H$\alpha$ Network \citep[GHN]{Steinegger.etal2000}, which is the \textit{adhoc} network of telescopes operated by different organizations. The data produced by these instruments are non-uniform in both spatial resolution and spectral bands, which may lead to a difference in the appearance of solar features.
Other notable examples are the U.S. Air Force Solar Optical Observing Network \citep[SOON]{Neidig.etal1998}, Kislovodsk High-Altitude Station of Pulkovo Observatory (KHASPO, Russia), Huairou Solar Observing Station (HSOS, P.R. China),  Hida Observatory (Kyoto University),
National Astronomical Observatory of Japan (NAOJ), Catania Astrophysical Observatory (CAO), Solar Survey Archive BASS2000 (SSABASS2000, France), H-$\alpha$ telescope at Big Bear Solar Observatory (HBBSO, California).
Early observations in H-$\alpha$ were taken on the photographic film. Currently, digitized (large) datasets include the Solar Digitization Project at New Jersey Institute of Technology (SDP, NJIT), Kodaikanal Solar Observatory (KSO, India), and NSO flare patrol telescope (NSO FPT). 
All ground-based observations are subject to local weather and instrument shutdown/repairs. Some digitized (scanned) imaging data are still missing digitization of metadata (e.g., time and date of observations), which could benefit tremendously from the application of pattern/text recognition and machine learning.} See Table~\ref{tab:datasources} for links to these data sources.

\begin{table}[tbph]
    \centering
    \begin{footnotesize}
    \begin{tabular}{|c|c|}
    \hline
        H-$\alpha$ Instrument & Link to data source  \\
        \hline
        Global H-$\alpha$ Network (GHN) & \url{http://www.bbso.njit.edu/Research/Halpha/}\\
        \hline
        U.S. Air Force SOON & \url{https://www.soonar.org/}\\
        \hline
        KHASPO, Russia & \url{http://en.solarstation.ru/}\\
        \hline
        HSOS, China & \url{https://sun10.bao.ac.cn/}\\
        \hline
        Hida Observatory & \url{https://www.hida.kyoto-u.ac.jp/SMART/}\\
        \hline
        NAOJ, Japan & \url{https://solarwww.mtk.nao.ac.jp/mitaka_solar/}\\ 
        \hline
        CAO & \url{http://ssa.oact.inaf.it/oact/image_archive.php}\\
        \hline
        SSABASS2000, France & \url{https://bass2000.obspm.fr/home.php}\\
        \hline
        HBBSO, California & \url{http://www.bbso.njit.edu/Research/FDHA/}\\
        \hline
        SDP, NJIT & \url{http://sfd.njit.edu/}\\
        \hline
        KSO, India & \url{https://kso.iiap.res.in/new}\\
        \hline
        NSO FPT & \url{https://nispdata.nso.edu/ftp/flare_patrol_h_alpha_sp/}\\
        \hline
    \end{tabular}
    \end{footnotesize}
    \caption{Links to data sources of instruments mentioned in Section~\ref{subsec:Halpha}.}
    \label{tab:datasources}
\end{table}

\subsection{GOES: Solar Flare List and X-ray Flux}
\label{subsec:GOES}

Solar flare events are recorded in the NOAA Geostationary Operational Environmental Satellites (GOES) flare list \citep{Garcia:1994}. The flare list is popularly used by the space weather community, while some recent works show that the list misses several major flare events~\citep{van2022solar}. The number of solar flare events is not massive enough to train large-scale machine-learning models with the amount of solar imaging data we have. For example, from 05/01/2010 to 06/20/2018, $12,012$ solar flares are listed with class, start, end, and peak intensity time of each event. There are five levels of flare events, A/B/C/M/X, ranging from the weakest to the strongest. Table~\ref{tab:strong_weak_flares_years} gives the number of flares of the B/C/M/X class recorded in the GOES data set. The A/B are considered weak flares, which do not impact the Earth much, whereas the M/X are considered strong flare events; thus, early detection is crucial. 

\begin{table}[ht]
\centering
\begin{footnotesize}
    \begin{tabular}{|c|cccccccccccccc|}
    \hline  & 2010 & 2011 & 2012 & 2013 & 2014 & 2015 & 2016 & 2017 & 2018 &	2019	& 2020	& 2021	& 2022 &	2023\\
    \hline
A	&0 &	1	&0	&0	&0	&0	&0	&0	&5	&0	&2&	0	&0&	0\\
B	&693	&629	&473	&467	&183	&446	&758	&632	&255	&188	&324	&1056	&22	&2\\
C	&156	& 996	&1115	&1192	&1622	&1274	&297	&230	&12	&32	& 76	&215	& 205	& 287	\\
M	&23	&106	& 124	& 97	& 194	& 128	& 15	& 37	& 0	&0	&0	&14	&21	&46\\
X	& 0	&8	&7	&12	&16	&2	&0	&4	&0	&0	&0	&2	&1	&1	\\
\hline
Total	& 872	&1740	&1719	&1768	&2015	&1850	&1070	&903	&272	& 220	&402	&1287	&249	&336\\
\hline
    \end{tabular}
    \end{footnotesize}
\caption{The number of flares of A/B/C/M/X classes recorded yearly from 2010 to 2023 in the GOES data set. }
    \label{tab:strong_weak_flares_years}
\end{table}

Figure~\ref{fig:goesxrayflux} shows the 1-minute GOES X-ray flux data updated in real-time from SWPC/NOAA while tracking solar flare events. The SWPC/NOAA definition of the start, peak, and end time of an X-ray event is as follows. ``The begin time of an X-ray event is defined as the first minute, in a sequence of 4 minutes, of steep monotonic increase in 0.1-0.8 nm flux.
The X-ray event maximum is taken as the minute of the peak X-ray flux.
The end time is when the flux level decays to a point halfway between the maximum flux and the pre-flare background level.'' In the duration of 04/27/2024-05/03/2024 (in Figure~\ref{fig:goesxrayflux}), there is one M4.4 flare occurring. 
\begin{figure}[tbph]
    \centering
    \includegraphics[width=0.9\textwidth]{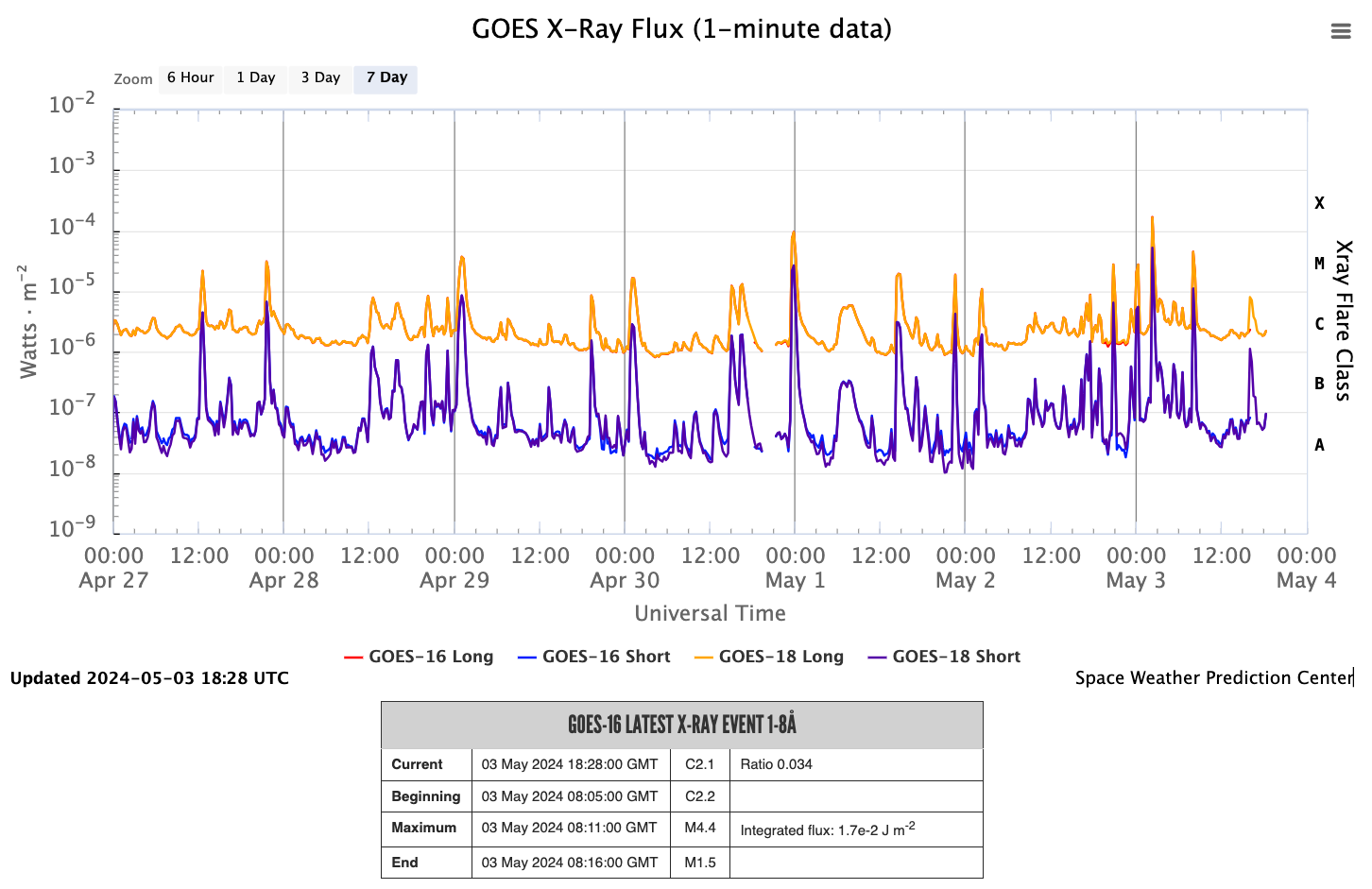}
    \caption{GOES X-ray flux record, downloaded from \url{https://www.swpc.noaa.gov/products/goes-x-ray-flux} on May 3, 2024. Raw data is also available for everyone to download.}
    \label{fig:goesxrayflux}
\end{figure}

\subsection{SDO: 3D Multi-channel Solar Imaging Data}
\label{subsec:SDO}
 Since its launch in 2010, NASA's Solar Dynamics Observatory (SDO; \citealt{pesnell2012}) has continuously observed solar activity, providing an extensive array of scientific data for heliophysics research. There are three instruments onboard: The Atmospheric Imaging Assembly (AIA; \citealt{Lemen:2012}) captures full-disk images of the Sun in high spatial (4096$\times$4096, pixel size of 0.6 arcseconds) and high temporal (12 seconds for EUV channels 94, 131, 171, 193, 211, 304, and 335 \AA, and 24 seconds for UV channels 1600 and 1700 \AA) resolution. The Helioseismic and Magnetic Imager (HMI; \citealt{Sherrer:2012}) captures visible wavelength filtergrams of the full Sun at 4096$\times$4096 resolution (pixel size of 0.5 arcsec). {Note that AIA has a slightly larger FOV (Field of View) ($\sim$41 arcmin) than HMI ($\sim$34 arcmin), which leads to the different pixel sizes mentioned above.} These filtergrams are then processed into various data products, including photospheric Dopplergrams, line-of-sight magnetograms, and vector magnetograms \citep{Hoeksema:2014a}. The EUV Variability Experiment (EVE; \citealt{woods2012}) measures the solar EUV spectral irradiance from 1 to 1050 \AA. Figure~\ref{fig:hmiaiadata} shows an example of the HMI and AIA full disc data. Tables~\ref{tab:SDO} and~\ref{tab:spectralbands} give more information on the coverage and properties of the SDO data.

\begin{figure}[tbph]
    \centering
    \includegraphics[width=0.9\textwidth]{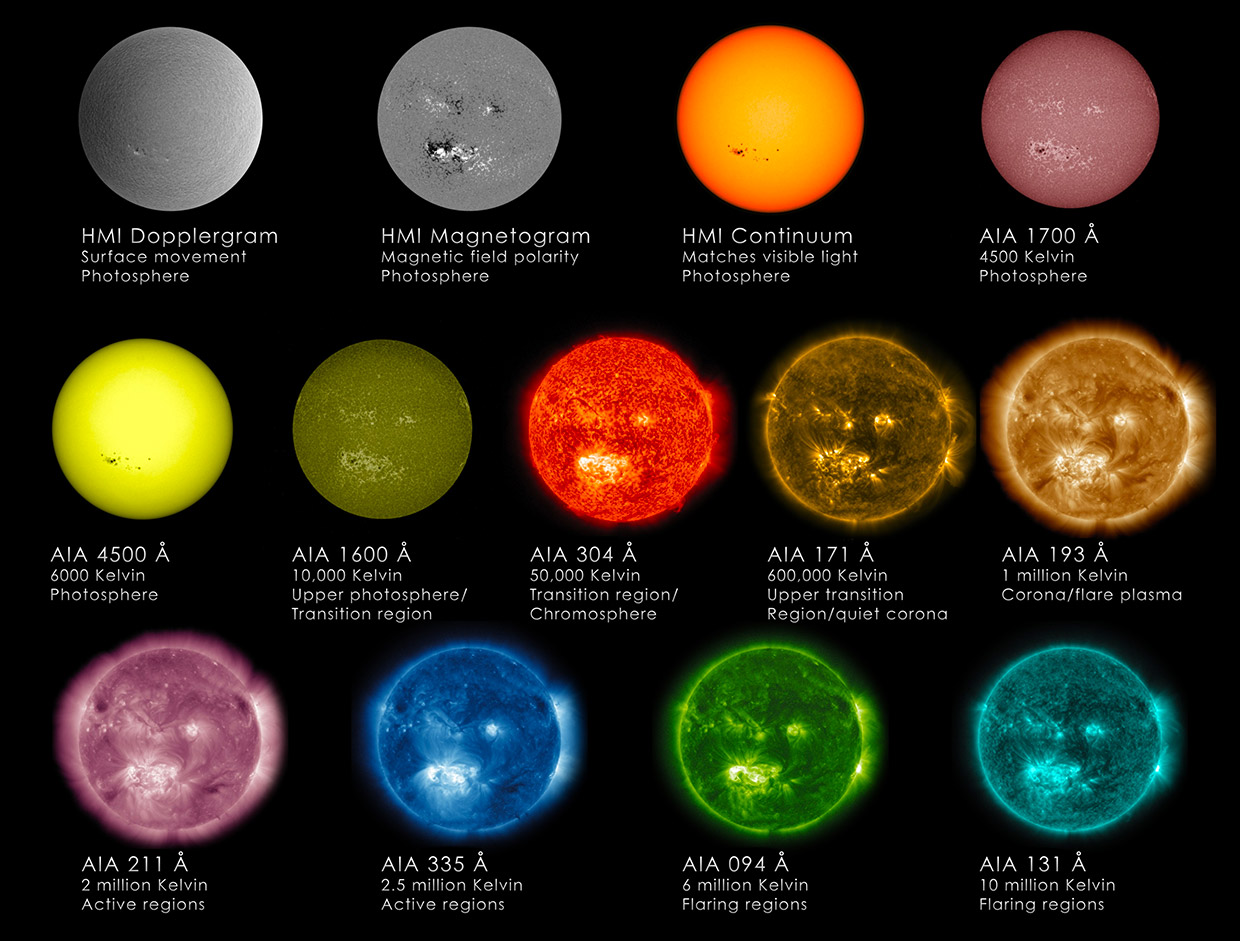}
    \caption{An example of the HMI and AIA data from \url{https://www.thesuntoday.org/sun/wavelengths/}}
    \label{fig:hmiaiadata}
\end{figure}

\begin{table}[ht]
\centering
\begin{footnotesize}
\begin{tabular}{|c|c|c|c|}
\hline
Instrument & Products & Cadence & Range \\
\hline
 HMI & LOS Magnetograms & 720 \& 45 secs & 2010.05.01 $\rightarrow$ \\
HMI &LOS Dopplegrams &720 \& 45 secs & 2010.05.01 $\rightarrow$\\
HMI & Continuum Intensity & 720 secs& 2010.05.01 $\rightarrow$\\
AIA &  Wavelengths 94,131,171,193, 211,304 and 335 \AA&  12 sec&  2010.05.13 $\rightarrow$\\
AIA &  Wavelengths 1600 and 1700 \AA &   24 sec&  2010.05.13 $\rightarrow$\\
AIA & Wavelength 4500 \AA & 1 hour & 2010.05.13 $\rightarrow$\\
\hline
\multicolumn{4}{|c|}{Derived products: Vector field, SHARP, Synoptic maps, Helioseismology }\\
\hline
\end{tabular}
\end{footnotesize}
\caption{Suite of SDO Instruments and products.} 
\label{tab:SDO} 
\end{table} 

\begin{table}[tbph]
    \centering
    \begin{footnotesize}
        
    \begin{tabular}{c|c|c|c|c}
    \hline
Band	& FWHM	 & Primary role	& Region of the  &	Typical Temperature
\\
& ($\Delta\lambda$, \AA)&ion(s) & Sun's atmosphere &(as $\log T[K]$)\\
\hline
6173 \AA &	75 m\AA	&HMI scans	&Intensity, velocity, and 	&3.7\\
&&	Fe i 6173& magnetic field of photosphere	& \\
\hline
4500 \AA &	500	&Continuum&	Photosphere&	3.7\\
1700 \AA&	200	&Continumm	&Temperature minimum, photosphere&	3.7\\
304 \AA&	12.7&	He ii	&Chromosphere, transition region	&4.7\\
1600 \AA&	200	&C iv, continuum	&Transition region, upper photosphere	&5.0\\
171 \AA	&4.7&	Fe ix	&Quiet corona, upper transition region	&5.8\\
193 \AA	&6.0&	Fe xii, xxiv	&Corona and hot flare plasma	&6.1, 7.3\\
211 \AA	&7.0&	Fe xiv	&Active region corona	&6.3\\
335 \AA	&16.5&	Fe xvi	&Active region corona	&6.4\\
94 \AA	&0.9&	Fe xviii&	Flaring regions	&6.8\\
131 \AA	&4.4&	Fe xx, xxiii&	Flaring regions	&7.0, 7.2\\
\hline
    \end{tabular}
    \end{footnotesize}
    \caption{HMI and AIA Spectral Bands Table directly replicated from \url{https://sdo.gsfc.nasa.gov/data/channels.php}.}
    \label{tab:spectralbands}
\end{table}

\paragraph{SDO Machine Learning Dataset (SDOML).} While SDO data are easily accessible, pre-processing these data for a scientific analysis often requires specialized heliophysics knowledge. To facilitate the SDO data usage, \citet{galvez2019} created a curated data set from the SDO mission in a format suitable for machine learning research. The SDOML Dataset is preprocessed from the original Level 1 data by down-sample HMI and AIA images from 4096$\times$4096 to 512$\times$512 pixels, removed $\rm{QUALITY} \ne 0$ observations, corrected for instrumental degradation over time, and applied exposure corrections. Both AIA and HMI data are spatially colocated and have identical angular resolutions {(pixel size of $\sim$4.8 arcsec)} 
, and that all instruments are chronosynchronous. The temporal resolution for AIA is 6 minutes. The temporal resolution for HMI vector magnetic field observations in Bx, By, and Bz components is 12 minutes. The EVE observations in 39 wavelengths from 2010-05-01 to 2014-05-26. The temporal resolution is 10 seconds. The dataset is permanently stored at the Stanford Digital Repository in .npz format. There is a recent update on this dataset, in which the entire dataset has been converted to a cloud-friendly Zarr format with complete FITS header information. Both the v1 and v2 versions of the dataset are now available through NASA HelioCloud on AWS (\url{https://registry.opendata.aws/sdoml-fdl/}).

\paragraph{Active Regions (AR) and SHARP Parameters.} Previous work has established that solar eruptions are all associated with highly nonpotential magnetic fields that store the necessary free energy. The most energetic flares come from very localized intense kilogauss fields of Active Regions (ARs) \cite{Forbes:2000a, Schrijver:2009a}.  At any time, these ARs occupy a small area of the solar surface. The magnetic complexity of active regions and their evolution often serve as the precursors of their flaring potential. Many CMEs are associated with the activation and subsequent eruption of the chromospheric filaments. Thus, the current applications of machine learning in solar and heliospheric physics are concentrating on identifying these solar phenomena.} The full disk HMI magnetogram data are subdivided into much smaller AR patches, where several of them often occur simultaneously; see Figure~\ref{fig:ARs_infulldisc} for the active region patches. Therefore, when handling AR-only data, the number of subdivided images that we have is another order of magnitude larger. In contrast, the amount of total data is significantly reduced (from 6 TB to 413 GB).

From each SHARP patch, scalar parameters (called SHARP parameters) are calculated to 
capture the zeroth order structure and complexity of the magnetic field. As discussed in \citet{Leka:2003} and \citet{Bobra:2014a}, these parameters are designed to assess the flaring potential of active regions. They are thus strongly representative of the total free energy of the magnetic field. These whole-active-region magnetic quantities can be effectively used as predictors of flares and also CMEs \citep[cf.][]{Falconer:2001, Falconer:2002, Falconer:2003, Falconer:2006, Leka:2003, Schrijver:2007, bobra2015solar}.

\paragraph{An Example Processing and Application of SDO Data.} In our data pre-processing pipeline~\citep{chen2019identifying}, the GOES flare list is matched to the Space-weather HMI Active Region Patch (SHARP) vector field data patches provided by the Joint Science Operations Center (JSOC) website. The SHARP patches contain 2-D photospheric maps of $3$ orthogonal magnetic field components observed by the Helioseismic and Magnetic Imager (HMI) on Solar Dynamics Observatory (SDO) with 1.0 arcsecond spatial resolution (4096x4096 pixel images) and time cadence of 12 minutes \citep{Hoeksema:2014a,Bobra:2014a}. This gives us $3\times 120$ high-resolution images per day from the three channels, which amounts to 
$>1$ million high resolution images over the $8$ year period that we work on. Due to the rotation of the Sun, an active region cannot be seen clearly (within 68 degrees of the central meridian to avoid foreshortening) for more than approximately 250 hours, which corresponds to $1250$ time frames at a time.

\begin{figure}[ht]
    \centering
    \includegraphics[width=0.9\textwidth]{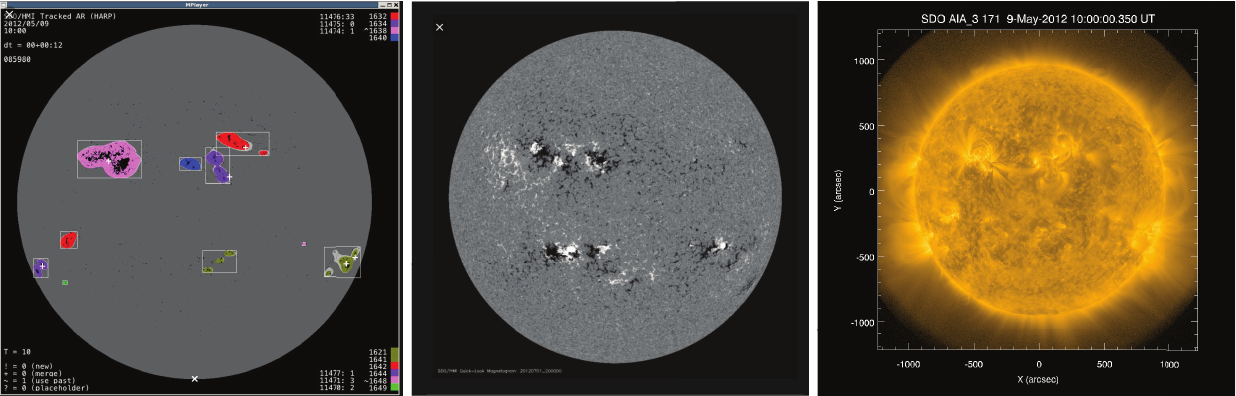}
    \caption{Solar Dynamics Observatory data used for flare prediction. Left panel: HMI active regions patches (HARPs) are highlighted in colored contours enclosed in boxes. Center panel: HMI magnetogram data are shown on a grayscale, where intense active regions are shown as opposite magnetic polarities, appearing in black and white. Right panel: AIA extreme ultraviolet image taken in the 171 Angstrom band.  Note the approximate co-location of the intense magnetic fields and the enhanced emission.}
    \label{fig:ARs_infulldisc}
\end{figure}

\subsection{SOHO and GONG: 2D Solar Imaging Data}
\label{subsec:SOHOGONG}

2D Line-of-sight (LOS) imaging data has been widely adopted for training solar flare forecasting models in the literature, as reflected by Table~\ref{tab:papersflares}. The most popular LOS images come from SOHO and GONG. The latter is supposed to be more suitable for operational use, whereas the former has been used more heavily by researchers for machine learning models~\citep[e.g.,][]{ji2022solar,sun2022predicting,guastavino2023operational}. Table~\ref{tab:SoHO_intext} shows the list of SoHO instruments. 

\begin{table}[ht]
\centering
\begin{footnotesize}
\begin{tabular}{|l|l|l|l|l|l|}
\hline
\multirow[c]{2}{*}{Instrument} & 
\multirow[c]{2}{*}{Observation} & 
\multirow[c]{2}{*}{Observed Region} & 
\multirow[c]{2}{*}{$\lambda($\AA$)$} & 
\multirow[c]{2}{*}{\minitab[c]{Cadence \\ (min)}} & 
\multirow[c]{2}{*}{Date Range}\\
 & & & &  &  \\
\hline
MDI & LOS Mag. Fld. & Full Disk & $6768$ (Ni I) & $\sim 96$ & 1996.05.01 - 2011.04.12 \\
\hline
EIT  & Intensity & Full Disk & $171$ (Fe IX/X) & $\sim 360$ & 1996.01.01 $\rightarrow$ \\
EIT & Intensity & Full Disk & $195$ (Fe XII) & $\sim 12$ & 1996.01.01 $\rightarrow$ \\
EIT & Intensity & Full Disk & $284$ (Fe XV) & $\sim 360$ & 1996.01.01 $\rightarrow$ \\
EIT  & Intensity & Full Disk & $304$ (He II) & $\sim 360$ & 1996.01.01 $\rightarrow$ \\
\hline
LASCO-C2 & Intensity & Corona ($1.5 - 6~R_s$) & Visible & $\sim 20$ & 1995.12.08 $\rightarrow$ \\
LASCO-C3 & Intensity & Corona  ($3.5 - 30~R_s$) & Visible & $\sim 20$ & 1995.12.08 $\rightarrow$ \\
\hline
\end{tabular}
\end{footnotesize}
\caption{\label{tab:SoHO_intext}Suite of SoHO Instruments. LOS Mag. Fld. denotes the line-of-sight magnetic field, $\lambda($\AA$)$ is wavelength measured in angstroms, and $R_s$ is the Sun's radius. LASCO C1 ($1.1 - 3~R_s$) is not included in this work since it was only operational until Aug. 9, 2000.} 
\end{table} 

Figure~\ref{fig:gongmag} shows the GONG magnetogram data from the NSO website, accessed on 3 May 2024. 
\begin{figure}[tbph]
    \centering
    \includegraphics[width=0.8\textwidth]{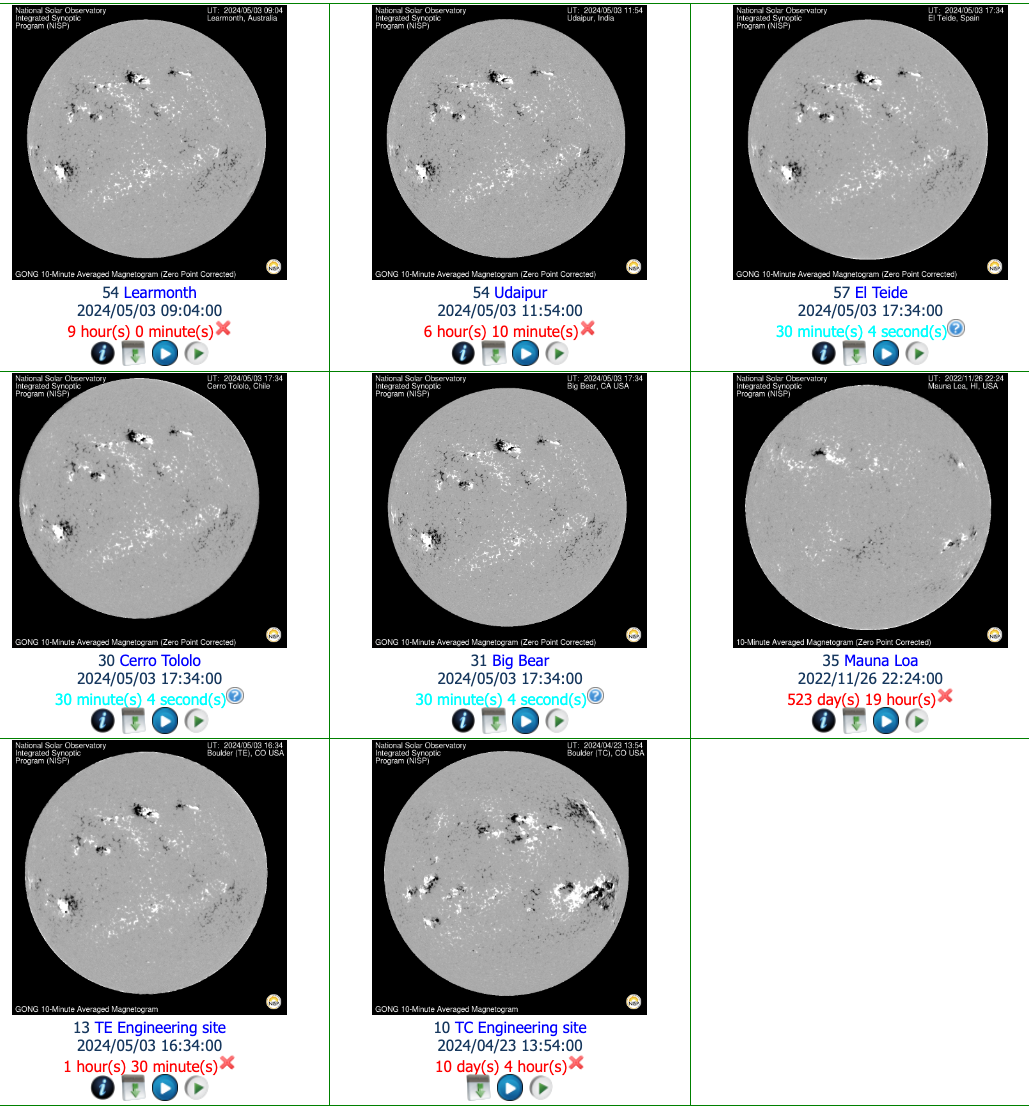}
    \caption{\footnotesize{GONG magnetogram data downloaded from the National Solar Observatory (NSO) data. 
    Summary of GONG LOS magnetograms from 3 May 2024 showing the latest observations from six network sites located at (upper panel, left to right) Learmonth Solar Observatory, Australia; Udaipur Solar Observatory, India; Teide Observatory, Canary Islands and (middle panel) the Cerro Tololo Inter-American Observatory, Chile; Big Bear Solar Observatory, California; and Mauna Loa Observatory, Hawai'i. Lower panels show magnetograms taken at two engineering sites in Boulder, Colorado. Notations below each image indicate the date and time of observations. The last line under each image shows the time since the last image was taken. For this line, red text is used for sites that are not being observed (due to weather or a night). The text in green shows the sites currently observing, and sites that are observing but have not returned a recent image (e.g., due to clouds) are shown in cyan.  GONG site at Mauna Loa Observatory was down since 26 Nov. 2022 after the volcanic eruption, and the lava flows shut down the operations. Hence, the image of the Sun from Mauna Loa is outdated and looks very different from the other sites. Similarly, observations from one of the engineering sites are also outdated. Magnetic fields are shown as black/white patches, corresponding to negative/positive polarity fields. 
    This summary page is available at \url{https://gong2.nso.edu/products/tableView/table.php?configFile=configs/averageMagnetogram10min.cfg}. Due to the dynamic nature of this page, the images for a different date and time will be shown. All images are oriented with the solar North up and East to the left.
    }}
    \label{fig:gongmag}
\end{figure}

\subsection{Existing Data Products and Software}


In the space weather literature, the data products are open-access and available to researchers to download. However, preprocessing of the raw data requires quite a bit of domain science and machine learning expertise. In response to this, research has been done on pre-processed ML-ready data for solar flare predictions. \citet{angryk2020multivariate} publishes ``a comprehensive, multivariate time series (MVTS) dataset extracted from solar photospheric vector magnetograms in Spaceweather HMI Active Region Patch (SHARP) series.''. The dataset covers 4,098 MVTS data collections from active regions occurring between May 2010 and December 2018, includes 51 flare-predictive parameters, and integrates more than 10,000 flare reports. However, this data product is not based on raw solar imaging but on summary statistics of solar patches, despite its nice properties of data preprocessing. 

{Table~\ref{tab:linkstodata} links to commonly used code to download, process, and visualize the solar imaging data products described in the subsequent paragraphs in this section.}

\begin{table}[tbph]
    \centering
{    \begin{tabular}{|c|c|}
    \hline
       Data  & Link  \\
       \hline
         Sunpy & \url{https://www.sunpy.org/}\\
         DRMS & \url{https://docs.sunpy.org/projects/drms/en/stable/}\\
Aiapy & \url{https://aiapy.readthedocs.io/en/stable/}\\
SDOML &  \url{https://github.com/spaceml-org/helionb-sdoml}\\
SDAC & \footnotesize{\url{https://hpde.io/NASA/NumericalData/SDO/AIA/NWRA/AARP/PT12S}}\\
\hline
    \end{tabular}}
    \caption{Links to sample code for solar data access. \texttt{Sunpy} is for reading and downloading general solar data. For example, \url{https://docs.sunpy.org/en/stable/generated/api/sunpy.net.dataretriever.GONGClient.html} provides access to the Magnetogram products of NSO-GONG synoptic Maps. \texttt{Aiapy} is for reading AIA data and is now one of the affiliated packages in \texttt{Sunpy}. Another sample code repository for SDOML is available at \url{https://gitlab.com/frontierdevelopmentlab/living-with-our-star/expanding-sdo-capabilities}.}
    \label{tab:linkstodata}
\end{table}

\paragraph{JSOC and SunPy.} Stanford University's Joint Science Operation Center (JSOC) stores data from SoHO MDI, SDO HMI and AIA, and various other solar instruments. The SunPy-affiliated package DRMS enables querying these images~\citep{Glogowski2019, sunpy20}.  These individual image products from JSOC are at the same processing level and are supplied in a Flexible Image Transport System (FITS) format containing only scalar values.
The NASA Solar Data Analysis Center's (SDAC) Virtual Solar Observatory\citep{vso} (VSO) tool enables data queries from a number of individual data providers.


\paragraph{AIA Active Region Patches (AARP) Database.} In contrast to the SDOML that down-sampled the full-disk images in space and time to keep the dataset size manageable, the AARP database preserves the native spatial resolution of SDO/AIA well samples the temporal evolution of solar active regions to match the SDO/HMI HARP database \citep{dissauer2023, leka2023}. The current AARP database includes daily 7-hour samples of 13 minutes of images (centered hourly on ``*.48UT" from 15:48 to 21:48 UT) from June 2010 to December 2018. The database aims to capture both the short-term dynamics (72 seconds' temporal resolution in 13 minutes' sample data) and long-term changes (7 hours of evolution per day). The total size of the database is $\sim$9 TB. This database is available through the NASA Solar Data Analysis Center (SDAC). 

\paragraph{A New Data Product in Pipeline for Publication.} Another SDO dataset that will be used in the studies mentioned in Section 5 is based on a similar idea as AARP but tailored particularly for ML-based flare prediction studies (Jin et al. 2024, in prep). Starting from the existing HMI HARP product, the AIA observations~\citep{Lemen:2012,Boerner:2012a} in 8 channels (94, 131, 171, 193, 211, 304, 335, 1600~\AA) are processed for the same FOV (Field of View) and coordinate (i.e., CEA: cylindrical equal area). The dataset includes all {B-class, C-class, and M-class flares from 2010-2024}, starting 24 hours before the flare onset until flare peak time with a temporal resolution of 12 minutes. In addition, we derive the Differential Emission Measure (DEM) maps 1 hour before each flare, which could provide additional information for the ML model. The total size of the dataset is $\sim$9 TB.

In addition to these aforementioned data products, scientists have been actively working on building ML-ready data products that will be made publicly available in the future. This is exemplified by the NASA call for proposals ``Heliophysics Artificial Intelligence / Machine Learning Ready Data'' (NNH23ZDA001N-HARD, ROSES-23 B.16 HARD) in 2023 and 2024. 

\section{Current Work with Solar Imaging Data}

In this section, we describe the current literature on the progress of machine learning approaches for handling solar data, including those that work with summary statistics instead of raw solar imaging data. 
Table~\ref{tab:papersflares} we provide a brief summary of the (partial) literature on solar flare predictions, including a partial list of representative papers published in recent years on this topic. {In solar flare predictions literature, the majority of the work adopts standard machine learning models, as given in the table. The predicted quantity is either the binary indicator of a strong or weak flare or the (logarithm of) peak flux intensity of a flare event. The predictors are either summary statistics (e.g., the SHARP parameters) or raw solar imaging data. Cross entropy loss and mean squared error loss are the most commonly adopted loss functions in the solar flare prediction literature. In binary classification, i.e., strong/weak flare prediction, because the samples are highly imbalanced (the number of strong flares is much smaller than the number of weaker flares), researchers have been using the HSS (Heidke skill score) and TSS (true skill score) as the metric for performance evaluation of prediction models~\citep[see e.g.,][]{chen2019identifying}. It is still an ongoing endeavor for researchers to obtain fair comparisons of the performances of machine learning models for solar flare forecasting. The difficulty lies primarily in the fact that different researchers/papers process and prepare the training, validation and testing data in their own ways, resulting in a lack of fair comparisons of results despite using the same metrics. For example, splitting data randomly versus splitting by non-overlapping years or splitting by active regions can result in significantly different results, as noted in our previous studies~\citep{wang2020predicting}. Furthermore, due to the scarcity of samples of strong solar flares, there are also non-negligible variations among different sample splits.}

In the following few subsections, we will describe in more detail the common practices that researchers in space weather have adopted to approach the solar flare prediction problem, focusing mainly on our previous work on solar flare forecasting as an example of utilizing solar data for prediction models.

\begin{table}[tbph]
    \centering
    \begin{footnotesize}
            \begin{tabular}{|c|c|c|}
    \hline
        Method & Data or Purpose & Paper(s)  \\
        \hline
        Deep Neural Network & Time Series of features & \citet{nishizuka2018deep,nishizuka2021operational}\\
        \hline
        AlexNet, GoogLeNet, DenseNet & Full-disk magnetograms & \citet{park2018application}\\
        \hline
        LSTM & X-ray flux GOES & \citet{yi2020forecast}\\
        \hline
        Linear models & HMI + GOES & \citet{anastasiadis2017predicting}\\
        \hline
        LSTM  & HMI + GOES & \citet{chen2019identifying,wang2020predicting}\\
        \hline
        LSTM & SHARP & \citet{liu2019predicting}\\
        \hline
        Mixed LSTM & HMI + GOES & \citet{jiao2020solar}\\
        \hline
        MLP, SVM, RF & NRT SHARP& \citet{florios2018forecasting}\\
        \hline
        FLARECAST (multiple) & R2O: SHARP+GOES & \citet{georgoulis2021flare}\\
        \hline
        \citet{florios2018forecasting} & \citet{angryk2020multivariate} & \citet{ji2020all}\\
        \hline
        Discriminant Analysis & DAFFS, operational & \citet{leka2018nwra}\\
        \hline
        Review paper & Operational flare forecasting & \citet{leka2019comparison}\\
        \hline
        LASSO+Fuzzy C-Means & SWPC Data &\citet{benvenuto2018hybrid}\\
        \hline
        CNN, DNN, bi-LSTM & Magnetogram+SHARP & \citet{tang2021solar}\\
        \hline
        Deep learning & LOS Magnetograms & \citet{huang2018deep,li2022knowledge}\\
        \hline
        SVM & SHARP & \citet{bobra2015solar}\\
        \hline
        Random forest & SWAN-SF & \citet{hostetter2019understanding}\\
        \hline
        Linear classifier & HMI+AIA 2010-2014 & \citet{jonas2018flare}\\
        \hline
        Multiple & SWAN-SF & \citet{ji2022solar} \\
        \hline
        Tree-based methods & DSD + SRS & \citet{cinto2020framework,cinto2020solar}\\
        \hline
        Video DNN & LOS images & \citet{guastavino2023operational}\\
        \hline
        Regression & HMI LOS + GOES X-ray & \citet{muranushi2015ufcorin}\\
        \hline
        Tensor-GPST & HMI+AIA images & \citet{sun2023tensor}\\
        \hline
        CNN & LOS of HMI & \citet{zheng2019solar}\\
        \hline
        CNN & Full-disc LOS & \citet{pandey2023explaining}\\
        \hline
        AlexNet, VGG16, and ResNet34 & Near-limb flares & \citet{pandey2023unveiling}\\
         \hline
    \end{tabular}
        \end{footnotesize}
    \caption{A list of representative papers published in recent years on solar flare predictions with machine learning approaches, together with the adopted method and data used. Acronyms: CNN (convolutional neural network), MLP (multi-layer perception), SVM (support vector machine), RF (random forest), DNN (deep neural network), LSTM (long short term memory), NRT (near real-time), R2O (research-to-operation), DAFFS (Discriminant Analysis Flare Forecasting System), SRS (Sunspot Region Summary), DSD (Daily Solar Data, NOAA/SWPC), NOAA (National Oceanic and Atmospheric Administration), SWPC (space weather prediction center), Tensor GPST (Tensor Gaussian Process with Spatial Transformation), SWAN-SF (Space-Weather ANalytics for Solar Flares \citep{DVNEBCFKM2020}). }
    \label{tab:papersflares}
\end{table}

\subsection{Feature Engineering Prior to Forecasting}

The most popular approach that researchers have adopted when handling solar imaging data is to perform feature engineering~\citep[e.g.,][]{jonas2018flare}, based on either physics or machine learning, prior to forecasting models. This efficiently reduces the dimensionality of the input space, which results in less over-fitting, cutting the data volume significantly and reducing the computational burden for prediction model training. {In \citet{stenning2013morphological}, morphological image analysis~\citep{soille1999morphological} techniques are adopted for solar images. It is shown that through the morphological image analysis, scientifically meaningful and interpretable numerical features can be extracted from high-throughput solar images for downstream classification and prediction tasks.} In \citet{chen2019identifying}, an autoencoder network is trained on HMI 3D magnetic field data, and the extracted features from the autoencoder are used to forecast strong solar flares. It is shown that the autoencoder can ``reconstruct'' the 3D magnetogram with a very small root mean squared error, but the extracted features do not outperform the known list of physics parameters (SHARPs~\citep{bobra2014helioseismic}) in terms of solar flare forecasting. A closer examination of the results shows that the reconstructed imaging data looks much smoother than the original observations. In the solar flare mechanism, the local information around the ``polarity inversion line'' is very important for flare forecasting~\citep{schrijver2007characteristic}. Therefore, black-box feature engineering may not be the best option to extract important features for strong solar eruptions from massive solar images with only a few hundred strong solar flare events.

In \citet{sun2021improved}, we investigate new features on top of the SHARP parameters for the flare classification task. The first set of features is derived from persistence homology in topological data analysis, following the idea in \citet{deshmukh2021shape}. We extend the scope of HMI images from just the $B_r$ {component} to multiple SHARP {parameter} maps when conducting the analysis and pay specific attention to the polarity inversion line region (PIL). The second set of features comes from spatial statistics concepts. The Ripley's K function analyzes the spatial clustering/dispersion patterns of pixels with high {$B_r$}. The Variogram analyzes the spatial variation of the {$B_r$} flux at various distance scales. Both sets of features summarize some information regarding the spatial distribution of SHARP parameters, which adds additional information to the feature set that SHARP parameters themselves cannot provide. We demonstrate how the new features can improve the skills of the prediction model and also show that new features, especially Ripley's K functions, have great discriminating power. See Figure~\ref{fig:4flare_example} for an illustration.

\begin{figure}[tbph]
    \centering
    \includegraphics[width=0.9\textwidth]{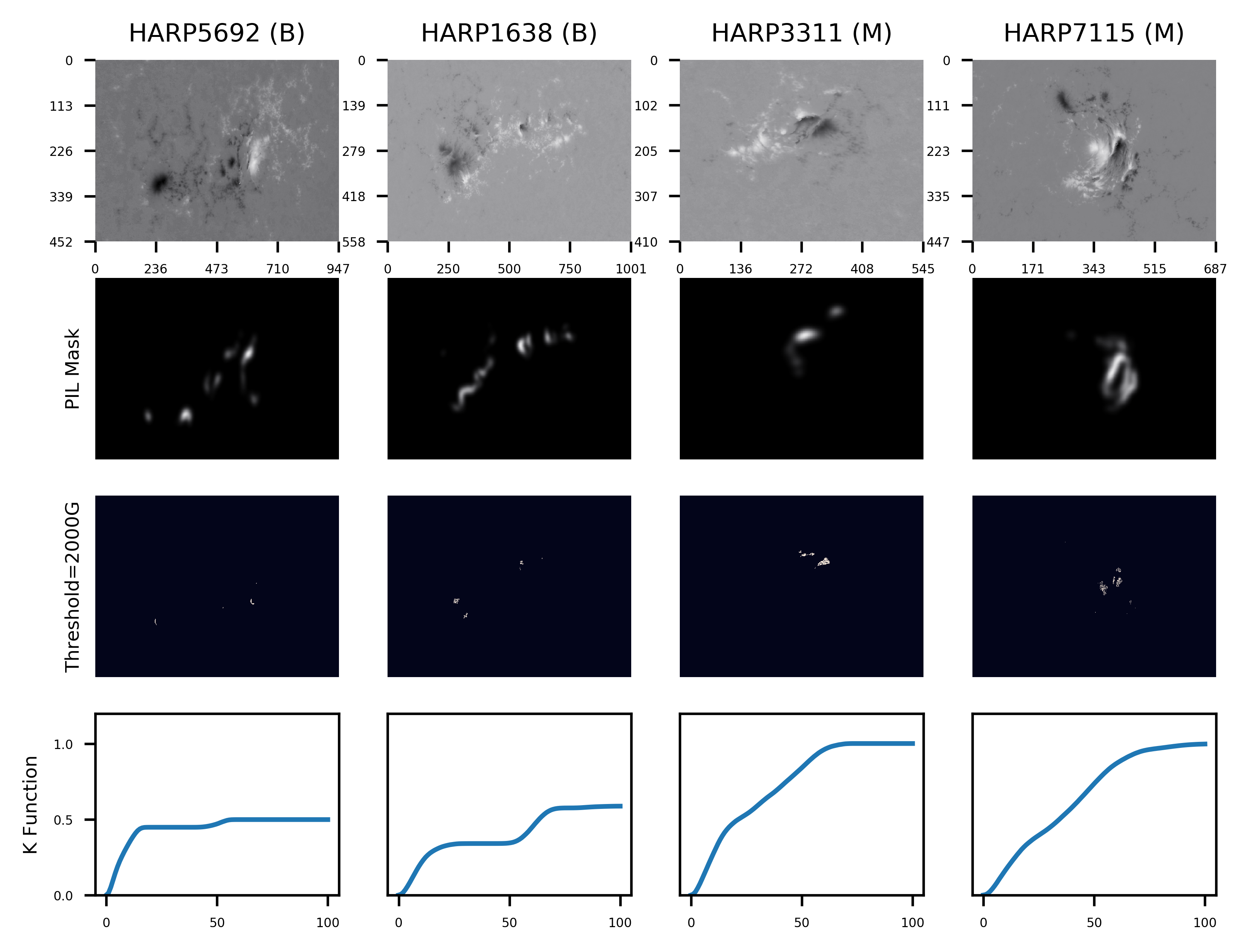}
    \caption{{Four flare examples (columns 1-4): B$6.1$ from HARP 5692 peaked at 04:36, Jun 26, 2015; B$5.3$ from HARP 1638 peaked at 02:23, May 09, 2012; M$1.0$ from HARP 3311 peaked at 19:53, Oct 26, 2013; M$1.0$ from HARP 7115 peaked at 03:51, Sept 05, 2017. The four rows correspond to their $B_r$ values, PIL masks, point clouds with $B_r>2000$ G within each PIL region, and Ripley's K functions, respectively. The K function differs, in terms of level and shape, between two M flares and two B flares. The main reason is that there are only scattered small clusters of high-$B_r$ regions for the B flares. On the contrary, M flares have a sub-region full of high-$B_r$ pixels.}}
    \label{fig:4flare_example}
\end{figure}
 
The findings of a strong correlation between the $B_r$ spatial distribution and the flare productivity in \citet{sun2021improved} shows an inherent connection between the free energy buildup and release in solar flares that is related to the clustering and proximity of flux to the PIL, which has been established earlier by \cite{Falconer:2003} and \cite{Schrijver:2005, Schrijver:2007} who respectively found the gradient and proximity of the magnetic flux (line-of-sight component) with respect to the PIL to be strongly correlated with flares and coronal mass ejections. This study shows the success of combining physics knowledge and classical statistics methodology (spatial statistics and topological data analysis) when constructing features for solar flare forecasting. The authors envision future work of this type can also benefit more and more data-driven approaches for space weather forecasting problems beyond solar flare forecasting.

\subsection{Prediction Model with Multi-channel Segmented Solar Images}

The structured information of the high dimensional multi-way tensor data makes the analysis an intriguing but challenging topic for statisticians and practitioners, especially in the context of physical sciences. In the solar flares prediction problem, the observations come as a time series of the 3-way tensor of active regions (see section~\ref{subsec:SDO}) of the Sun that produces flares: 3-D magnetosphere maps from the Helioseismic and Magnetic Imager and multi-channel temperature maps from the Atmospheric Imaging Assembly~\citep{bobra2014helioseismic}. Figure~\ref{fig:flare-example-original} shows eight AIA channels, one HMI channel, and one Polarity Inversion Line (PIL) for an active region on the Sun that produced an M-class (strong) flare. The data size is $377\times 744 \times 10$ for this one-time point, and we have data from 2010 till now every 12 minutes.

\begin{figure}[htb]
    \centering
    \includegraphics[width=0.95\textwidth]{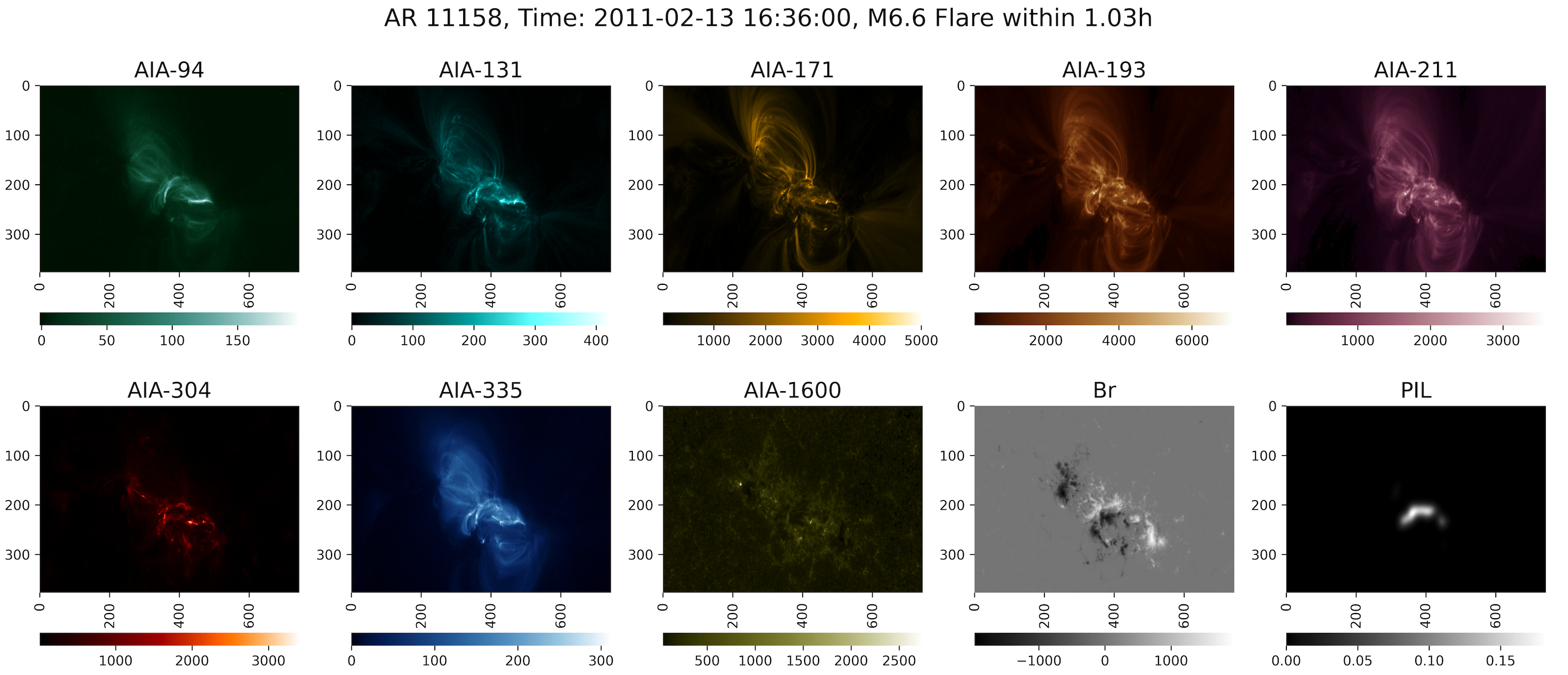}
    \caption{M-class Flare Example for Active Region (AR) No.11158, recorded at 16:36:00 (UT) of Feb 13, 2021. The flare intensity is $6.6\times 10^{-5} \mbox{W/m}^{2}$ and peaked at 17:38:00 (UT) of the same day.}
    \label{fig:flare-example-original}
\end{figure}

In our recent paper in ICML, \citet{sun2023tensor}, we expand the Tensor-GP model by integrating a dimensionality reduction technique, called \textit{tensor contraction}, with a Tensor-GP for a scalar-on-tensor regression task with multi-channel imaging data. We first estimate a latent, reduced-size tensor for each data tensor and then apply a multi-linear Tensor-GP on the latent tensor data for prediction. We introduce an anisotropic total-variation regularization when conducting the tensor contraction to obtain a sparse and smooth latent tensor. We then propose an alternating proximal gradient descent algorithm for estimation. We validate our approach via extensive simulation studies and apply it to the solar flare forecasting problem. More precisely, we consider a multi-channel imaging dataset $\{\mathcal{X}_{i}, y_{i}\}_{i=1}^{N}$, where {the multichannel solar images} $\mathcal{X}_{i} \in \mathbb{R}^{H\times W\times C}$ with $H, W, C$ as the height, width and number of channels, respectively; and {the logarithmic peak flare intensity} $y_{i}\in \mathbb{R}$. We use $\mathbf{X}_{i}^{(c)} \in \mathbb{R}^{H\times W}, c\in[C]$ to denote the $c^{\rm{th}}$ channel of $\mathcal{X}$. Our model is specified as
\begin{equation}\label{eq:Baseline-GP}
    y_i = f(g(\mathcal{X}_{i})) + \epsilon_{i}, \quad f(\cdot) \sim \mbox{GP}\left(m(\cdot), k(\cdot,\cdot)\right),
\end{equation}
with $\epsilon_{i} \sim \mathcal{N}\left(0, \sigma^{2}\right)$ being the idiosyncratic noise and $g(\mathcal{X}_{i}) = \mathcal{X}_{i} \times_{1} \mathbf{A} \times_{2} \mathbf{B} \times_{3} \mathbf{I}_{C}$ with $\mathbf{A} \in \mathbb{R}^{h\times H}, \mathbf{B} \in \mathbb{R}^{w\times W}$, and $\mathbf{A}$ and $\mathbf{B}$ reduce the dimension of each channel of $\mathcal{X}_{i}$ from $H\times W$ to $h\times w$. All channels share the same tensor contracting factors $\mathbf{A}$ and $\mathbf{B}$, which preserves the spatial consistency of different channels of the reduced-sized tensor $\mathcal{Z}$ for easier interpretation.

\subsection{Prediction Model with Full Disc Solar Images}


Despite the fact that it is known based on heliophysics knowledge, that energetic solar flares are localized in small patches of the Sun, named active regions (ARs, see section~\ref{subsec:SDO} for details and references), there are continuous efforts among machine learning and heliophysics community on constructing prediction models with full disc solar images, with the hope that the machine can \textit{figure out} the active regions in an automated fashion. The pros of this approach are that it is not restricted to \textit{established knowledge} and thus has the potential of identifying solar flares that are not localized in active regions if there exist any. The cons of this approach is that it needs to rely on the limited number of strong solar flares to construct very sparse features from very high-dimensional imaging data, which can result in over-fitting when training machine learning models. However, if adopted properly, the authors believe that the large volume of raw solar imaging data for both quiet time and flaring time can potentially be utilized to train a large black-box machine learning model with competitive performances. Whether or not this model will outperform those models informed by physics knowledge or supplemented by laws of plasma physics remains unknown. 

\citet{pandey2023explaining} uses CNN for ``hourly full-disk line-of-sight magnetogram images and employs a binary prediction mode to forecast $\geq$M-class flares that may occur within the following 24-h period''. The study shows that the full-disc analysis is aligned with precursors that occur in active regions. It is claimed that the trained model can learn shape and texture-based characteristics, even if in near-the-limb regions, where the majority of the previous literature does not consider~\citep[e.g.][]{chen2019identifying}. Furthermore, in a follow-up work by the same group, \citet{pandey2023unveiling}, focuses on the near-the-limb regions and explores multiple networks for strong flare forecasting. 

\section{Looking into the Future: Gaps and Opportunities}
\label{sec:future}

In summary, an extensive set of machine learning algorithms has been tried out to show promising results of data-driven solar flare forecasting. Other solar eruptions, such as the more severe solar energetic particles, have also been tested with data-driven approaches~\citep{kasapis2022interpretable,whitman2022review,kasapis2024forecasting}. However, the majority of the works directly convert/prepare the solar data into a classical binary classification problem, ignoring various levels of complications of the solar imaging data and solar flaring properties. We list a few of the major features of the solar imaging and flaring data here, thus pointing out challenges and opportunities for developments in statistical theory, methodology, and computational algorithms.

\subsection{Solar Cycle}
The Sun's activity follows approximately 11-year cycles, which is shown in Figure~\ref{fig:solarcycle}, represented by the sunspot numbers. During the peak of a solar cycle, called solar max, energetic solar eruptions are far more frequent than during the valley of a solar cycle, called solar minimum. \citet{qahwaji2007automatic} uses the sunspot groups and solar cycle data to forecast strong solar flares. \citet{wang2020predicting} examines the solar cycle dependency of solar flare forecasting with the LSTM method, showing statistically significant variation when different years of data are chosen for training and testing. This leads to issues when comparing different models in the literature. Furthermore, the high-quality observations, for example, the SDO data, only cover one solar cycle. This makes it hard to learn repeating patterns for solar eruptions from purely data-driven approaches. Statistical models, especially Bayesian statistical methods, will turn out to be efficient in incorporating such information, especially considering operational settings. Furthermore, it is important to know that the solar cycle is not a deterministic quantity (periodicity). It is shown in Figure~\ref{fig:solarcycle} that we are currently experiencing a much stronger solar cycle than projected (predicted by NASA). Therefore, accounting for the uncertainty in solar cycle forecasting, within the solar eruption forecasting, is also of vital importance. Figure~\ref{fig:xrayradio} shows the maximum F10.7 cm radio flux (typically three are recorded each day) against the proportion of minutely GOES X-ray fluxes at or above the C-class threshold of $10^{-6} W/m^2$. {The data ranges from 2004-10-28 to 2024-01-25.}

\begin{figure}[tbph]
    \centering
    \includegraphics[width=0.6\textwidth]{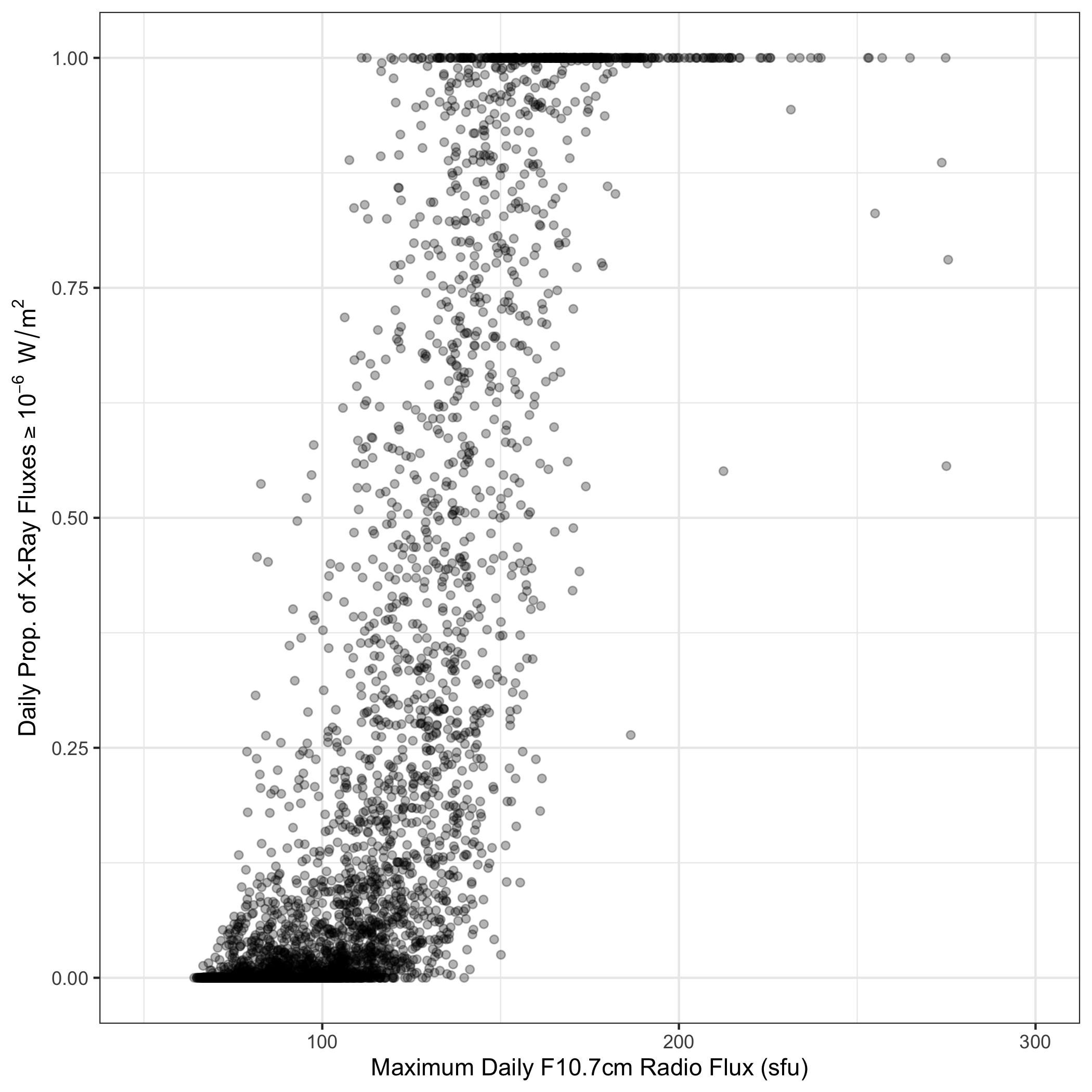}
    \caption{The maximum F10.7 cm radio flux (typically three are recorded each day) against the proportion of minutely GOES X-ray fluxes at or above the C-class threshold of $10^{-6} W/m^2$.}
    \label{fig:xrayradio}
\end{figure}

\subsection{Heterogeneity}

The heterogeneity of properties of active regions and flaring mechanisms, on top of the solar cycle dependence, contribute to the difficulties of predicting strong solar flare events. The different active regions have very different ``lifetimes'', very different sizes, and can erupt drastically different numbers of solar flares. The determination of the lifetime of active regions is also biased by the fact that direct observations are limited to the time when the region is located in the solar hemisphere facing Earth, and thus, the regions could emerge and/or disappear during the time they are located on ``farside" of the Sun.

Figure~\ref{fig:harpinfor} shows histograms of active region lifetime and pixelsizes for active regions from 05/01/2010 till 02/24/2024. Furthermore, Table~\ref{tab:num_AR_num_events} shows the number of active regions (ARs) corresponding to the specified number (1, 2, 3, 4, 5, and $>5$) of weak (B) and strong (M/X) flare events for each active region recorded in the GOES data set from 05/01/2010 till 06/20/2018. The active region information (i.e., size) is typically used as a predictor in flare forecasting. However, in a majority of the works in literature, ``personalized'' forecasting for each active region as it emerges and fades has not been thoroughly investigated. An earlier paper, \citet{wheatland2004bayesian}, proposed a Bayesian method that uses the flaring record of an active region to refine an initial prediction for the occurrence of a big flare during a subsequent period of time. This shows the initial success of solar flare forecasting when taking into account active region evolution. 

\begin{figure}[tbph]
    \centering
    \includegraphics[width=0.75\textwidth]{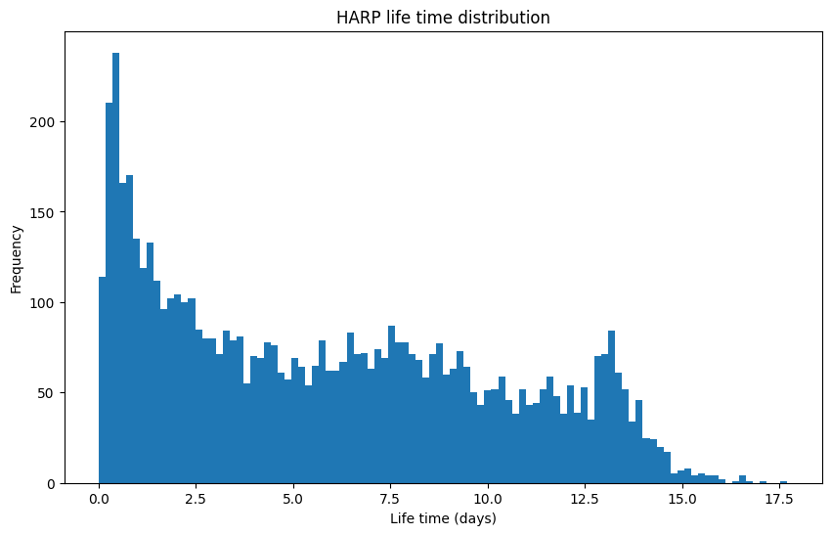}\\
    \includegraphics[width=0.75\textwidth]{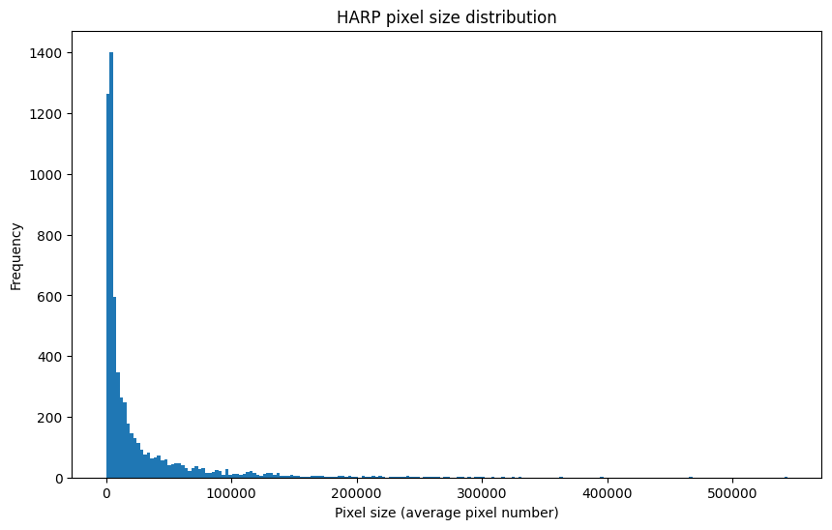}
    \caption{HARP lifetime and pixel sizes.}
    \label{fig:harpinfor}
\end{figure}

\begin{table}[tbph]
    \centering
    \begin{tabular}{|c|cccccc|}
    \hline
    Number of M/X Flares & 1 & 2 & 3 & 4 & 5 & $> 5$\\
    \hline
    Number of ARs & 60 & 31 & 13 & 10 & 7 & 29\\
    \hline
    \hline
    Number of B Flares & 1 & 2 & 3 & 4 & 5 & $> 5$\\
    \hline
    Number of ARs & 321 & 148 & 51 & 19 & 2 & 0\\
    \hline
    \end{tabular}
    \caption{Number of active regions (ARs) corresponding to the specified number (1, 2, 3, 4, 5, and $>5$) of weak (B) and strong (M/X) flare events for each active region {recorded in the GOES data set} from 05/01/2010 till 06/20/2018. }
    \label{tab:num_AR_num_events}
\end{table}

\subsection{Noisy and Missing Data} 

The missing data problem is rarely discussed in solar eruptions literature and has been handled straightforwardly, such as linear interpolation or removing missing segments. However, proper handling of missing data is important statistically, in general. There are several types of missing data in solar observations. Some can be ignored or simply interpolated (such as due to instrument failure or cloud passing by), and some cannot be ignored. There are missing patches of full solar images at certain time points, missing pixels for a continuous time range, and missing labels for solar images. Furthermore, the GOES solar flare list is also shown to miss major events~\citep{van2022solar}. Imputing missing data properly requires a thorough understanding of the missing data mechanism, e.g., missing at random, not-missing-at-random, and systematic missingness~\citep{little2019statistical}.

\subsection{Sample Sparsity}

Solar flare samples are very sparse, especially considering the strong flares ($\geq M$-class). Other solar eruptions, such as the solar energetic particles, are even sparser. What further complicates the issue is that in solar minimum, we have even sparser samples of solar eruptions, whereas, in solar maximum, consecutive strong solar eruptions might be highly correlated with each other. Training black-box machine learning algorithms, especially complicated neural network structures, relies on enough number of samples. There has been work on ``data augmentation''~\citep{shorten2019survey} in the machine learning literature of creating synthetic samples for training machine learning models, which has been applied in the context of solar flare forecasting~\citep{ji2022solar}. Furthermore, denoising diffusion models are becoming more and more popular nowadays for creating samples~\citep[e.g.][]{ho2020denoising,song2023solving} that represent the training data. These techniques must be scrutinized within the context of the operational solar eruptions section by the joint work of machine learning experts and operational scientists. 

\subsection{Computational Efficiency}

Despite the sample sparsity problem, the raw (most likely multi-channel) imaging data that is observed every few seconds or minutes spanning over the range of a decade or two can be large in volume (e.g., tens of terabytes). The computational efficiency of statistical and machine learning methods plays an important role in the successful implementations of the algorithms on solar imaging data. {As it currently stands in literature, to make models computationally feasible to fit, data with reduced temporal and/or spatial resolution (or even summary statistics, such as the SHARP parameters or extracted features, instead of raw multichannel solar imaging data) are adopted for training/validating/testing purposes in machine learning models. \citet{lai2021method} propose a divide-and-conquer method (with generalized fiducial inference) capable of handling massive datasets and providing uncertainty estimates. They analyzed the solar flare brightness with the proposed approach using the SDO/AIA data from \citet{schuh2013large}.}

\subsection{Uncertainty Quantification}

Properly quantifying the uncertainty for space weather monitoring and forecasting is essential for risk assessment and decision making~\citep{licata2022uncertainty}. Probabilistic prediction of solar eruptions should include not only the probability of observing a major eruption in a future time, as given in the majority of the papers in the literature but also an uncertainty estimate of the probability. Given a major eruption, the uncertainty estimate of the intensity of a solar eruption is also of importance. For example, the peak X-ray flux intensity of a strong solar flare can be an important predictor for a solar proton event~\citep{kasapis2022interpretable}. 

\subsection{Research-to-Operation (R2O)}

From SWPC/NASA, it is defined that ``Research-to-Operations-to-Research (R2O2R) refers to the cyclical process by which basic research endeavors (R), having been identified as having the potential for improving forecasting capabilities, are matured, in a targeted way, toward a formal operational implementation (O) and, once \textit{operationalized}, subsequent needs for refinements are conveyed back to the research community (R).'' Data-drive solar eruption forecasting is motivated by and will finally serve operational use. The research-to-operation takes far more validation and testing than ``a successfully trained model''. We need to handle real-time forecasting updates and model evolution over time before being able to put data-driven models into operation. This current solar cycle demonstrates quite intense Sun activities and thus can serve as an excellent test-bed for operational forecasting models.

\subsection{Other Space Weather Phenomena} Finally, we want to mention that many other relevant space weather phenomena can benefit from data-driven approaches, such as the geomagnetic index prediction~\citep{iong2022new,iong2024sparse,ren2020statistical} and the total electron content map reconstruction~\citep{sun2022matrix,sun2023complete} and prediction~\citep{sun2023matrix,wang2023forecast}. \citet{Camporeale:2019a} gives an excellent review of machine learning challenges in space weather from nowcasting and forecasting perspectives.

\section{Summary}

The authors hope this short introduction to solar imaging data can generate more interest in the statistics community to develop theory, methods, and computational algorithms to tackle the thorny data-analytic challenges presented by solar imaging data and, more broadly, space weather observations.







\section*{Disclosure Statement}

YC is supported by NSF DMS 2113397, NSF PHY 2027555, NASA 22-SWXC22\_2-0005, and 22-SWXC22\_2-0015. The authors report there are no competing interests to declare. The authors thank Victor Verma (Ph.D. student at the University of Michigan) for creating Figure~\ref{fig:xrayradio}, Ke Hu (Master's student at the University of Michigan) for creating Table~\ref{tab:strong_weak_flares_years} and Figure~\ref{fig:harpinfor}, and Enrico Camporeale (University of Colorado Boulder) for helpful discussions on solar data. 



\bibliographystyle{abbrvnat}
\bibliography{references}

\end{document}